\documentclass[prb,twocolumn,amssymb,amsmath,eqsecnum,%
floatfix,showpacs]{revtex4}
\usepackage{bm,dcolumn}
\newcommand{\vect}[1]{\mathbf{#1}}
\newcommand{\order}{O}

\begin{document}
\title{First-principles envelope-function theory for lattice-matched
       semiconductor heterostructures}
\author{Bradley A. Foreman}
\email{phbaf@ust.hk}
\affiliation{Department of Physics,
             Hong Kong University of Science and Technology,
             Clear Water Bay, Kowloon, Hong~Kong, China}

\begin{abstract}
In this paper a multi-band envelope-function Hamiltonian for
lattice-matched semiconductor heterostructures is derived from
first-principles self-consistent norm-conserving pseudopotentials.
The theory is applicable to isovalent or heterovalent heterostructures
with macroscopically neutral interfaces and no spontaneous bulk
polarization.  The key assumption---proved in earlier numerical
studies---is that the heterostructure can be treated as a weak
perturbation with respect to some periodic reference crystal, with the
nonlinear response small in comparison to the linear response.
Quadratic response theory is then used in conjunction with $\vect{k}
\cdot \vect{p}$ perturbation theory to develop a multi-band
effective-mass Hamiltonian (for slowly varying envelope functions) in
which all interface band-mixing effects are determined by the linear
response.  To within terms of the same order as the position
dependence of the effective mass, the quadratic response contributes
only a bulk band offset term and an interface dipole term, both of
which are diagonal in the effective-mass Hamiltonian.  The interface
band mixing is therefore described by a set of bulk-like parameters
modulated by a structure factor that determines the distribution of
atoms in the heterostructure.  The same linear parameters determine
the interface band-mixing Hamiltonian for slowly varying and
(sufficiently large) abrupt heterostructures of arbitrary shape and
orientation.  Long-range multipole Coulomb fields arise in quantum
wires or dots, but have no qualitative effect in two-dimensional
systems beyond a dipole contribution to the band offsets.  The method
of invariants is used to determine the explicit form of the
Hamiltonian for $\Gamma_6$ and $\Gamma_8$ states in semiconductors
with the zinc-blende structure, and for intervalley mixing of $\Gamma$
and $X$ electrons in (001) GaAs/AlAs heterostructures.
\end{abstract}


\maketitle

\section{Introduction}

\subsection{Background and motivation}

Envelope-function models continue to play a key role in the design and
interpretation of experiments on semiconductor heterostructures.  The
canonical ``envelope-function approximation,'' which by definition
makes use of only bulk effective-mass parameters and heterojunction
band offsets, has been very successful in explaining a wide range of
experiments. \cite{Bast88,Bast91} However, recent studies have
increasingly emphasized interface-related effects lying outside the
scope of conventional envelope-function theory, such as optical
\cite{KrVo96,KrVo98,Plat99} and electrical
\cite{TakVol00b,Klip02a,RosKai02} anisotropy, intervalley mixing,
\cite{Puls89,AlIv93,FWIK93,Klip99,Klip00,Klip01,GliKra00} and spin
polarization phenomena.
\cite{Gan01,Gan02a,Gan02b,Golub03,Gan03b,Belkov03,Gan03c,Gan04a,%
Sin04,Shen04,ZhangYang05,BernZhang05a} These effects are generated
(wholly or in part) by interface band-mixing terms in the
heterostructure Hamiltonian, such as valence-band mixing of light and
heavy holes, \cite{AlIv92,IvKaAl93,IvKaRo96} $\Gamma$--$X$ coupling,
\cite{AlIv93,FWIK93} and the conduction or valence band Rashba
coupling. \cite{Wink03}

A great deal of progress in the modeling of these effects can be made
on the basis of symmetry information alone.  The method of invariants,
which was originally developed for bulk semiconductors,
\cite{Lutt56,SuzHen74,BirPik74,TrRoRa79} has proved a powerful tool in
the study of heterostructures as well.
\cite{IvPi97,AlIv92,IvKaAl93,IvKaRo96,GliKra98a,GliKra98b,GliKra00,%
TakVol00b,Gan01,Gan02a,Gan02b,Golub03,Gan03b,Belkov03,Gan03c,Gan04a,%
Sin04,Shen04,ZhangYang05,BernZhang05a,Wink03}
The standard method of invariants uses symmetry information to
construct an explicit interface Hamiltonian, but the same information
(supplemented by hermiticity or current-conservation requirements) may
also be used to construct connection rules for the envelope functions
on opposite sides of the
interface. \cite{Laik92,deLeon98,Kisin93,Kisin94,KiGeLu98,%
TokTsiGor02,Rod02,Rod03} In either case, one obtains a
phenomenological model containing some interface parameters whose
value is determined by comparison with experiment.

Useful as this approach may be, it does not provide any information
about the magnitude of the interface parameters, some of which (for
reasons unrelated to symmetry) may happen to be zero or negligibly
small.  To obtain this information (along with a deeper understanding
of the physical origin of the interface phenomena), one must turn to a
more detailed microscopic model of the interface.  Thus, numerous
envelope-function models have been derived directly from the
microscopic potential energy; these include interface Hamiltonians
\cite{Leib75,Leib77,Burt88a,Burt92,Burt94c,Burt99,Young89,CuVH91,%
KaTi91,KaKr96a,KaKr96b,Fore93,Fore95a,Fore96,Fore98b,Fore01b,BuFo99,%
Klip98,TakVol97a,TakVol97b,TakVol99a,TakVol99b,TakVol00a,%
Cortez00a,Cortez01c} and connection rules
\cite{SmMa86,SmMa90,MaSm90a,Trze88b,AnWaAk89,CuVH92,CuVH93a,%
CuVH93b,AvSi93a,AvSi94,GrKaCh94} as well as numerical approaches based
on solving the $\vect{k} \cdot \vect{p}$ equations in momentum space.
\cite{WaZu96,YiRaz97}

However, all of the cited derivations are based on empirical
pseudopotentials, in which the potential energy is determined by the
choice of some specific model for the interface.  This leads to
ambiguity in the results, as different choices may yield conflicting
predictions as to which interface parameters are important,
\cite{Fore98b,Klip98,Klip99,Klip00,Klip01,TakVol00a} or whether
interface band-mixing effects are directly related to band offsets.
\cite{Fore99,Cortez00a,Cortez01c,MagZun00} Such conflicts can only be
resolved by deriving the envelope-function Hamiltonian from an
{\em ab initio} self-consistent potential, as suggested by Sham and
Lu.  \cite{ShamLu89} The purpose of this paper is to present such a
derivation and examine its implications for interface band-mixing
effects.  Numerical applications of the theory are not considered
here.  A preliminary account of these results has been presented
elsewhere. \cite{Fore04}

\subsection{Basic assumptions and limitations}

The current ``standard model'' for condensed-matter physics is based
on density-functional theory in the local density approximation (LDA).
\cite{HoKo64,KoSh65,Payne92} However, it is well known that this
ground-state formalism does not accurately predict energy gaps in
semiconductors.  Thus, the present state-of-the-art in band-structure
theory involves calculations of the electron self-energy in the
shielded-interaction \cite{BaymKad61,Baym62} or $GW$ approximation.
\cite{Hedin65,HedLund69,Aulbur00,Onida02} This field is not yet fully
mature, as apparent early successes
\cite{HybLou86,GodSchSham88,ZhuLou91} have been questioned in light of
recent developments.  \cite{UmBoHa98,Eguil98,Arn00,Eguil02,Leb03}
Nevertheless, any future refinements in approximation techniques for
calculating the quasiparticle band structure will continue to be based
on the self-energy in Dyson's equation.  For this reason, the present
work relies only upon general properties of the self-energy operator,
not upon any specific approximation to this quantity.

For simplicity, however, this paper assumes that the ionic cores are
described in terms of norm-conserving pseudopotentials.
\cite{HamSchChi79,Klein80,BacSch82,BaHaSc82,TroMar91} The projector
augmented wave (PAW) method of Bl\"{o}chl, \cite{Blochl94,Arn00,Leb03}
which does not involve this approximation, has become increasingly
popular in recent years.  A generalization of the present theory to
incorporate the PAW formalism should be possible, but to avoid undue
complexity it is not considered here.  In addition, this paper
considers only lattice-matched systems (with no bulk or interface
strain).

The key assumption used in the present work is that the
heterostructure is a small perturbation \cite{AndBalCar78} with
respect to some periodic virtual reference crystal (such as
Al$_{0.5}$Ga$_{0.5}$As for a GaAs/AlAs heterostructure).  This
assumption has been verified within LDA for both isovalent and
heterovalent systems, including GaAs/\-AlAs,
\cite{ResBarBal89,BaReBaPe89} Ge/\-GaAs, \cite{BaReBa88,BaReBaPe89}
In$_{0.53}$Ga$_{0.47}$As/\-InP, \cite{BaReBaPe89,PBBR90}
GaAs/\-Si/\-AlAs and GaAs/\-Ge/\-AlAs, \cite{PBRB91,BaPeReBa92} Si/Ge,
\cite{ColResBar91,PerBar94} and
InAs/\-GaSb. \cite{PeMoBaMo96,MonPer96} Within the framework of
pseudopotential theory, one would expect it to be valid for other
similar heterostructures also.  This assumption makes it possible to
describe the self-consistent heterostructure perturbation in terms of
nonlinear response theory, with the linear response predominant and
the quadratic response providing a weak correction.
\cite{ResBarBal89,BaReBa88,BaReBaPe89,PBBR90,PBRB91,%
ColResBar91,BaPeReBa92,PerBar94,PeMoBaMo96,MonPer96}

Such an approach has been used by Sham \cite{Sham66} to derive an
effective-mass equation for shallow impurity states in semiconductors.
Sham's work was extended in the preceding paper \cite{Fore05a} to
obtain expressions for the self-energy at small values of the crystal
momentum in lattice-matched heterostructures described by
spin-dependent nonlocal pseudopotentials.  These expressions are used
here to construct a multi-band effective-mass theory for
lattice-matched heterostructures.

Although the basic formalism developed here is quite general, to keep
the paper to a reasonable length it is necessary to impose some
restrictions on the material systems that are treated in detail.
(This allows the power series in Sec.\ \ref{sec:power_series} to be
terminated at a reasonably low order.)  Since only lattice-matched
systems are considered here, it is assumed that the symmetry of the
reference crystal does not support a spontaneous polarization, because
that would generate macroscopic electric fields and piezoelectric
strain fields inconsistent with the lattice-matching assumption.
Therefore, the present theory is not directly applicable to wurtzite
materials.

In addition, it is assumed that the interfaces in a heterovalent
system such as Ge/GaAs are macroscopically neutral,
\cite{BaReBaPe89,Har78} so that the atoms can be grouped together into
neutral clusters (of fractional atoms) that carry a dipole moment only
at the interface. \cite{Fore05a,note:heterovalent} Thus, the main
application of the present theory would be to isovalent or
heterovalent heterostructures of semiconductors with the zinc-blende
or diamond structure.  However, the only examples treated explicitly
here are from isovalent systems.

Lattice mismatch could not be included in this theory without
fundamental changes to account for atomic relaxation.  Nevertheless,
the present results provide a solid foundation for subsequent
extensions of the theory to include this effect.  Other restrictions,
such as macroscopic neutrality and the lack of bulk dipole terms,
could be lifted merely by extending some power series expansions to
higher order (although, depending on the accuracy that is desired,
\cite{note:heterovalent} this may require the inclusion of certain
nonanalytic terms neglected in Ref.\ \onlinecite{Fore05a}).

The fact that (within the pseudopotential approximation) a typical
heterostructure is a weak per\-tur\-ba\-tion
\cite{AndBalCar78,ResBarBal89,BaReBa88,BaReBaPe89,PBBR90,PBRB91,%
ColResBar91,BaPeReBa92,PerBar94,PeMoBaMo96,MonPer96} makes possible
the existence of energy eigenstates $| \psi \rangle$ whose wave
function $\langle n \vect{k} | \psi \rangle$ in the Luttinger-Kohn
representation \cite{LuKo55} $|n \vect{k} \rangle$ of the reference
crystal is negligible outside a small region in $\vect{k}$ space near
the high-symmetry points of the Brillouin zone.
\cite{Xia89,WaFrZu97,WaZu97,WaZu99} These eigenstates are just the
low-energy excitations that are of greatest interest experimentally.
Their existence, which has been verified by extensive numerical work
on empirical pseudopotential models,
\cite{Xia89,WaFrZu97,WaZu97,WaZu99} is precisely the condition needed
for the validity of an effective-mass approximation for slowly varying
envelope functions $F_n (\vect{x})$,
\cite{LuKo55,Leib75,Leib77,Burt88a,Burt92,Burt94c,Burt99} since $F_n
(\vect{k}) \equiv \langle n \vect{k} | \psi \rangle$.  The existence
of slowly varying envelopes
\cite{Burt92,Burt94c,Xia89,WaFrZu97,WaZu97,WaZu99} provides the
foundation for the effective-mass theory derived here.

The present approach, based on quadratic response theory,
\cite{Fore05a} fits well with the $\vect{k} \cdot \vect{p}$
perturbation formalism of Leibler,
\cite{Leib75,Leib77,TakVol97a,TakVol97b,TakVol99a,TakVol99b,TakVol00a}
in which the heterostructure perturbation is treated as small in
comparison to the energy separating the bands of interest from other
remote bands.  Leibler's theory is used here to develop a multi-band
effective-mass theory that includes all terms of the same order as the
position dependence of the effective mass.  This includes cubic and
quartic dispersion terms in the Hamiltonian of the reference crystal,
as well as the leading contributions from the quadratic response.  As
shown in Sec.\ \ref{sec:error}, higher-order terms cannot consistently
be described in terms of local differential
equations. \cite{TakVol97a,TakVol97b,TakVol99a,TakVol99b,TakVol00a}

\subsection{Summary of key results}

\label{sec:summary}

In this paper, it is shown that the dominant interface band-mixing
terms are those arising from the linear response.  Indeed, the linear
response is the {\em only} contribution to band mixing that fits
within the framework of the perturbation theory defined above, and the
only contribution that can consistently be described in terms of local
differential equations.  To this level of accuracy, the quadratic
response contributes only a bulk band offset term and an interface
dipole term, neither of which produces any band mixing in the
effective-mass Hamiltonian.

This represents a major simplification, since it implies that the
interface band-mixing terms in the Hamiltonian are just a
superposition of parameters derived from the linear response to
individual ionic perturbations (or neutral cluster perturbations
\cite{Fore05a,note:heterovalent} in the case of heterovalent
substitutions).  These parameters are calculated once and for all for
a given material system; they then appear as coefficients in front of
structure factors describing the distribution of atoms in the
heterostructure.  The Hamiltonian contains spatial derivatives of (and
differences between) these structure factors, which generate
$\delta$-like functions at an interface.  The atomic distribution
functions change for heterostructures of different shape (wells,
wires, or dots) and orientation, but their coefficients do not.  Thus,
a single set of interface parameters governs the band mixing for any
type of heterostructure \cite{note:any} in a given material system.

If desired, one can express these linear parameters as a {\em
difference} between the properties of the various {\em bulk} materials
that make up the heterostructure.  (A similar result was obtained for
linear band offsets in Refs.\
\onlinecite{ResBarBal89,BaReBa88,BaReBaPe89,PBBR90,PBRB91,%
ColResBar91,BaPeReBa92,PerBar94,PeMoBaMo96,MonPer96}.)  However, in a
no-common-atom system such as InAs/GaSb, this must include the
(lattice-matched) ``interface'' materials GaAs and InSb.  Also, these
bulk-like linear response parameters (which in principle do not
require a supercell calculation \cite{Bar01}) cannot be determined
from experiments on bulk semiconductors.

For some terms in the linear response (derived from the analytic part
of the self-energy\cite{Fore05a}), the interface Hamiltonian itself
has the form of a macroscopic average of the Dirac $\delta$ function
or its derivatives.  For the remaining terms (derived from the
nonanalytic part of the self-energy\cite{Fore05a}), the effective
charge density has this localized form, but the potential energy is
not as well localized, having the form of a long-range multipole
potential.  However, in a quantum well or any other heterostructure
with two-dimensional translation symmetry, all terms are well
localized except for the interface dipole terms (which merely modify
the band offsets).

The band-mixing Hamiltonian has the same general form for slowly
graded structures (within the virtual crystal approximation) and
abrupt heterojunctions, the only difference being the rate of change
of the structure factor.  That is, in contrast to previous theories
based on model potentials or empirical pseudopotentials, no new
band-mixing parameters appear at an abrupt junction.  Indeed, the
localized interface terms in the Hamiltonian derived here are
qualitatively identical to the Hamiltonian derived by Leibler
\cite{Leib75,Leib77} for slowly graded heterostructures.

These results shed light on recent suggestions
\cite{Klip98,Klip99,Klip00,Klip01,TakVol00a} that the
$\Gamma_1$--$X_{1z}$ and $X_{1x}$--$X_{1y}$ intervalley mixing
potentials at an ideal (001) GaAs/AlAs heterojunction should be
proportional to $\delta(z)$.  It is shown here that such mixing arises
only from the quadratic response and is therefore negligible in
comparison to $\Gamma_1$--$X_{3z}$, $X_{1z}$--$X_{3z}$, and
$X_{3x}$--$X_{3y}$ mixing.  There is, however, a linear contribution
to the $\Gamma_1$--$X_{1z}$ and $X_{1x}$--$X_{1y}$ mixing that is
proportional to $\delta'(z)$, as well as a linear $\delta(z)$ mixing
for nonideal interfaces or when spin-orbit coupling is included.
These contributions may help to explain the experimental observations
in Refs.\ \onlinecite{Klip99,Klip00,Klip01}.

\subsection{Outline of the paper}

The paper begins in Sec.\ \ref{sec:error} with a discussion of which
terms are to be retained in $\vect{k} \cdot \vect{p}$ perturbation
theory, based upon a review of important recent work by Takhtamirov
and Volkov. \cite{TakVol97a,TakVol97b,TakVol99a,TakVol99b,TakVol00a}
The main results of the quadratic response theory developed in the
preceding paper \cite{Fore05a} are presented (in modified form) in
Sec.\ \ref{sec:qrt}.  The basic envelope-function formalism is
developed in Sec.\ \ref{sec:envelope}, while Sec.\
\ref{sec:power_series} describes the power series expansions that are
used to obtain approximate expressions valid for slowly varying
envelopes.  Perturbation theory is used to eliminate the coupling to
``remote'' bands in Sec.\ \ref{sec:perturbation}, yielding the basic
expression for the multi-band envelope-function Hamiltonian.
Modifications to the Hamiltonian that are necessary if one wishes to
describe the material parameters of an abrupt heterojunction as
piecewise constant are discussed in Sec.\ \ref{sec:stepfunction}.
Symmetry properties of the Hamiltonian are discussed in Sec.\
\ref{sec:symmetry}, where explicit matrix representations of the
material properties are given for semiconductors with the zinc-blende
structure.  Finally, the significance of the results and their
relation to previous work in the literature are discussed in Sec.\
\ref{sec:discussion}.

\section{Order of terms included}

\label{sec:error}

Takhtamirov and Volkov
\cite{TakVol97a,TakVol97b,TakVol99a,TakVol99b,TakVol00a} have recently
demonstrated (using an instructive analogy to the leading relativistic
corrections to the Schr\"odinger equation) that most derivations of
heterostructure effective-mass equations in the literature (including
those of the present author) do not consistently include all
perturbative corrections of the same order.  This section presents a
review and discussion of their results, with the objective of
establishing which terms are to be retained in subsequent perturbative
approximations.

The first case to be considered is a wide-gap system satisfying
\begin{equation}
   \Delta \bar{V} \ll \bar{E}_{\text{g}} , \label{eq:DV}
\end{equation}
where $\Delta \bar{V}$ is a typical heterojunction band offset, and
$\bar{E}_{\text{g}}$ is a typical energy gap (for the virtual bulk
reference crystal) separating the band in question from all other
``remote'' bands.  It is assumed that in the reference crystal this
band is describable by an effective-mass equation with effective mass
$m^*$.  This will be the case if the $\vect{k} \cdot \vect{p}$
interaction with remote bands
\begin{equation}
   \frac{\hbar \bar{k} \bar{p}}{m} \equiv (\bar{k} \lambda)
   \bar{E}_{\text{g}} \label{eq:lambda}
\end{equation}
can be treated as a small perturbation.  Here $m$ is the free-electron
mass, $\bar{k} \sim 2\pi/L$ is a typical envelope function wavenumber
for a quantum well of width $L$, $\bar{p} \sim 2\pi\hbar/a$ is a
typical interband momentum matrix element (where $a$ is the lattice
constant), and it is assumed that $L \gg a$.  Equation
(\ref{eq:lambda}) defines a length parameter $\lambda = \hbar \bar{p}
/ m \bar{E}_{\text{g}}$, which (if the free-electron contribution to
$m^*$ is negligible) may also be written as $\lambda \approx \hbar (2
m^* \bar{E}_{\text{g}})^{-1/2}$.  For GaAs, both expressions for
$\lambda$ give $\lambda \approx 6$ \AA, thus $\lambda \sim a$ and
$\bar{k} \lambda \ll 1$ in a wide quantum well.  For the remainder of
this paper the parameters $\lambda$ and $a$ are used interchangeably,
although in general error estimates should be based on the larger of
the two quantities.

In general one is interested in cases where the kinetic energy is
comparable to the band offset:
\begin{equation}
   \frac{\hbar^2 \bar{k}^2}{2 m^*} \sim \Delta \bar{V} \sim (\bar{k}
   \lambda)^2 \bar{E}_{\text{g}} .
\end{equation}
This is the order of terms included in ordinary effective-mass theory,
\cite{LuKo55} which may be compared with the nonrelativistic
(Schr\"odinger) approximation to the Dirac equation.  The position
dependence of the effective mass can be calculated by treating $\Delta
\bar{V} / \bar{E}_{\text{g}}$ as a perturbation, \cite{Leib75,Leib77}
yielding a correction of order
\begin{equation}
   \frac{\hbar^2 \bar{k}^2}{2 m^*} \left( \frac{\Delta
   \bar{V}}{\bar{E}_{\text{g}}} \right) \sim \frac{(\Delta
   \bar{V})^2}{\bar{E}_{\text{g}}} \sim (\bar{k} \lambda)^2 \Delta
   \bar{V} , \label{eq:pos_dep_mass}
\end{equation}
which is of the same order as the nonparabolic $k^4$ terms in the
kinetic energy of the reference crystal.  Hence, these terms
(analogous to the relativistic mass corrections to the Schr\"odinger
equation) must be included if one is to retain all terms of the same
order as the position dependence of the effective mass.

As will be seen below, the interface band-mixing terms proportional to
a $\delta$ function are of order
\begin{equation}
   A \langle \delta(z) \rangle \sim (\bar{k} a) \Delta \bar{V} ,
   \label{eq:d0}
\end{equation}
while those proportional to the derivative of a $\delta$ function are
of order
\begin{equation}
   B \langle \delta'(z) \rangle \sim (\bar{k} a)^2 \Delta \bar{V} ,
   \label{eq:d1}
\end{equation}
in which $A$ and $B$ are constants.  Hence, the latter terms are also
comparable to the position dependence of the effective mass.

For a step-function discontinuity $\Delta V$ in the potential energy,
the envelope function $F$ has a discontinuity in its second derivative
given by
\begin{equation}
   \frac{\Delta F''}{F} = \frac{2 m^*}{\hbar^2} \Delta V ,
\end{equation}
where $F \sim L^{-1/2}$.  This gives rise to an asymptotic behavior in
$k$ space of $F(k) \sim (\Delta F'') k^{-3}$.  But in the exact
Luttinger-Kohn envelope-function representation, \cite{LuKo55} the
envelope functions are limited to wave vectors inside the first
Brillouin zone.  Hence, the use of a local differential effective-mass
equation, which gives rise to nonvanishing $F(k)$ outside the
Brillouin zone, generates an error in the kinetic energy of order
\begin{equation}
   (\bar{k} a)^3 \left( \frac{\hbar^2 \bar{k}^2}{2 m^*} \right) \sim
   (\bar{k} a)^3 \Delta \bar{V} . \label{eq:local_error}
\end{equation}
Therefore, in the local approximation, the accuracy is limited by Eq.\
(\ref{eq:local_error}), and contributions beyond the level of
(\ref{eq:pos_dep_mass}) and (\ref{eq:d1}) should for consistency be
omitted.

Of course, this is not a fundamental limitation of the
envelope-function method, as one can always choose to work in
$\vect{k}$ space (which is a common choice for numerical work
\cite{BarGer91,GerHenBar93,WiRo93}).  However, there is another
similar source of error that arises from anticrossings of the bands
explicitly included in the envelope-function model with those treated
as remote perturbations.  This occurs, for example, in the $\Gamma_6$
conduction band of GaAs at about a third of the distance to the
Brillouin zone boundary in the $\Delta$ direction.  Beyond such an
anticrossing the model is no longer valid and an error of order
(\ref{eq:local_error}) arises even in a $\vect{k}$-space formalism.
To eliminate such errors one must enlarge the Hamiltonian by treating
these remote bands explicitly, thus obtaining a full-zone $\vect{k}
\cdot \vect{p}$ model. \cite{CarPol66,PoHiCa66,WaZu96,MaSm90a}

Such a multi-band envelope-function model is also needed for medium-gap
($\Delta \bar{V} \sim E_{\text{g}}^{(0)}$) and narrow-gap
($E_{\text{g}}^{(0)} \ll \Delta \bar{V}$) systems, where
$E_{\text{g}}^{(0)}$ is the energy gap of the reference crystal.  The
perturbative approach based on Eq.\ (\ref{eq:DV}) can still be used,
provided that $\bar{E}_{\text{g}}$ refers to the energy gap between
the bands of interest and those treated as remote perturbations.
Takhtamirov and Volkov \cite{TakVol00a} have considered the extreme
narrow-gap limit for the case in which the dispersion is dominated by
linear-$k$ terms for all energy ranges of interest, and showed that in
this case the error generated by the local approximation is of order
$(\bar{k} a)^2 \Delta \bar{V}$.  In such cases the terms
(\ref{eq:pos_dep_mass}) and (\ref{eq:d1}) should be omitted for
consistency.  However, in practical situations one is more likely to
encounter cases in which $\Delta \bar{V} \sim E_{\text{g}}^{(0)}$ and
the $k^2$ dispersion terms are comparable to the linear-$k$ terms.
Thus, in this paper all terms of order $(\bar{k} a)^2 \Delta \bar{V}$
are retained, while those at the level of the local approximation
(\ref{eq:local_error}) are omitted.

It should be noted that a fully self-consistent perturbation scheme is
not always desirable.  It would of course be impossible to achieve an
accurate numerical prediction for the fine structure of the hydrogen
atom without including the relativistic-mass correction and the Darwin
term in addition to the spin-orbit splitting.  However, for
qualitative considerations one is often interested primarily in
symmetry-breaking effects, in which case the former two contributions
may justifiably be omitted.  Likewise, although the present work
retains all terms of order $(\bar{k} a)^2 \Delta \bar{V}$, certain of
these (such as the $k^4$ bulk dispersion terms) may possibly be
omitted for applications in which the primary focus is on
symmetry-breaking interface effects.  [As an example, Takhtamirov and
Volkov have proposed a model\cite{TakVol00b} based on certain terms of
order $(\bar{k} a)^3 \Delta \bar{V}$, while neglecting larger terms of
order $(\bar{k} a) \Delta \bar{V}$ and $(\bar{k} a)^2 \Delta
\bar{V}$.] However, the neglect of these terms can only be justified
for specific individual applications and should not be presumed to
hold in general.

The discussion here has focused on wide quantum wells, but this should
not be taken to imply that effective-mass theory is inapplicable in
other cases.  For example, in a narrow quantum well,
\cite{Burt92,TakVol99b} an effective-mass approximation can be
developed along the same lines as for shallow
impurities. \cite{LuKo55,Sham66} In this case, however, the above
estimates of the interface terms and local approximation are no longer
valid; see Ref.\ \onlinecite{TakVol99b} for further details.

The remainder of this paper uses atomic units with $\hbar = m = e =
1$.

\section{Quadratic response theory}

\label{sec:qrt}

\subsection{Basic definitions}

The problem of interest is the Dyson eigenvalue equation (see Ref.\
\onlinecite{Fore05a} and references therein)
\begin{equation}
   - \tfrac12 \nabla^2 \psi(\vect{x}, \omega) + \int V(\vect{x},
   \vect{x}', \omega) \psi(\vect{x}', \omega) d^3 \! x' =
   E(\omega) \psi(\vect{x}, \omega) , \label{eq:schr_x}
\end{equation}
in which $\omega$ is a complex energy parameter, $E$ is the complex
eigenvalue, and the self-consistent potential $V$ is the sum of a
fixed norm-conserving ionic pseudopotential $V_{\text{ion}}$ and the
nonhermitian self-energy operator $\Sigma$:
\begin{equation}
   V(\vect{x}, \vect{x}', \omega) = V_{\text{ion}}(\vect{x},
   \vect{x}') + \Sigma(\vect{x}, \vect{x}', \omega) .
\end{equation}
For notational simplicity, the $\omega$ dependence will be suppressed
in most of the equations that follow.  In Eq.\ (\ref{eq:schr_x}),
$\psi$ and $V$ are spinors, with $V$ having the form
\begin{equation}
   V = \openone \, V_{\text{sc}} + \bm{\sigma} \cdot \vect{V} ,
   \label{eq:spinorV}
\end{equation}
in which $\openone$ is the $2 \times 2$ spinor unit matrix,
$V_{\text{sc}}$ is the scalar relativistic part of $V$, $\bm{\sigma}$
is the Pauli matrix, and $\vect{V}$ accounts for spin-orbit
coupling. \cite{HybLou86b,Surh91,HeFoNe93,Theur01} Note that Eq.\
(\ref{eq:schr_x}) incorporates all relativistic corrections of order
$Z^2 \alpha^2$ (where $Z$ is the atomic number and $\alpha$ is the
fine-structure constant), but neglects terms of order $\alpha^2$, such
as spin-orbit coupling outside the atomic
cores. \cite{Klein80,BacSch82} 

In a heterostructure, it is convenient to partition the ionic
pseudopotential as
\begin{equation}
   V_{\text{ion}} (\vect{x}, \vect{x}') = V_{\text{ion}}^{(0)}
   (\vect{x}, \vect{x}') + \Delta V_{\text{ion}} (\vect{x}, \vect{x}')
   ,
\end{equation}
where $V_{\text{ion}}^{(0)}$ is the ionic pseudopotential of some
periodic reference crystal (which may be a virtual crystal).  This is
defined as
\begin{equation}
   V_{\text{ion}}^{(0)} (\vect{x}, \vect{x}') = \sum_{a, j, \vect{R}}
   f^{aj} v^{a}_{\text{ion}} (\vect{x} - \vect{R}_{j}, \vect{x}' -
   \vect{R}_{j}) , \label{eq:V_ion_0}
\end{equation}
in which $v^{a}_{\text{ion}} (\vect{x}, \vect{x}')$ is the ionic
pseudopotential for atomic species $a$, $f^{aj}$ is the fractional
weight associated with atom \cite{note:atom} $a$ on site $j$ at
position $\bm{\tau}_{j}$ in the unit cell of the reference crystal,
and $\vect{R}_{j} = \vect{R} + \bm{\tau}_{j}$, where $\vect{R}$ is a
Bravais lattice vector of the reference crystal.  $f^{aj}$ must
satisfy
\begin{equation}
   0 \le f^{aj} \le 1 , \qquad \sum_{a} f^{aj} = 1 ,
\end{equation}
although the former constraint need not be strictly enforced.

The term $\Delta V_{\text{ion}}$ is the perturbation due to the
heterostructure:
\begin{equation}
   \Delta V_{\text{ion}} (\vect{x}, \vect{x}') = \sum_{a, j, \vect{R}}
   \theta^{aj}_{\vect{R}} v^{a}_{\text{ion}} (\vect{x} - \vect{R}_{j},
   \vect{x}' - \vect{R}_{j}) .
\end{equation}
Here $\theta^{aj}_{\vect{R}}$ is the change in fractional weight
(relative to the reference crystal) of atom $a$ at position
$\vect{R}_{j}$ in the heterostructure, which must satisfy
\begin{equation}
   0 \le f^{aj} + \theta^{aj}_{\vect{R}} \le 1 , \qquad \sum_{a}
   \theta^{aj}_{\vect{R}} = 0 . \label{eq:theta_constraint}
\end{equation}
The constraint (\ref{eq:theta_constraint}) permits one to rewrite
$\Delta V_{\text{ion}}$ as
\begin{equation}
   \Delta V_{\text{ion}} (\vect{x}, \vect{x}') = \sideset{}{'}
   \sum_{\alpha, \vect{R}} \theta^{\alpha}_{\vect{R}} \Delta
   v^{\alpha}_{\text{ion}} (\vect{x} - \vect{R}_{\alpha}, \vect{x}' -
   \vect{R}_{\alpha}) , \label{eq:Delta_V_ion}
\end{equation}
in which
\begin{equation}
   \Delta v^{\alpha}_{\text{ion}} (\vect{x}, \vect{x}') \equiv
   v^{a}_{\text{ion}} (\vect{x}, \vect{x}') -
   v^{\bar{a}_j}_{\text{ion}} (\vect{x}, \vect{x}')
   . \label{eq:Delta_v_ion}
\end{equation}
Here $\alpha = (a, j)$ is a composite index, $\bar{a}_j$ is the label
of some given atom on site $j$, and the prime on the summation symbol
indicates that the values $\bar{\alpha} = ( \bar{a}_j, j )$ are
excluded.  To simplify the interpretation of (\ref{eq:Delta_V_ion}),
the atom $\bar{a}_j$ is chosen to be the same as the (virtual) atom on
site $j$ in the reference crystal.

In this paper the self-consistent potential $V (\vect{x}, \vect{x}')$
is treated using nonlinear response theory.  \cite{Fore05a} The
fundamental assumption is that $V$ can be expressed as a power series
in the variables $\theta^{\alpha}_{\vect{R}}$:
\begin{equation}
   V (\vect{x}, \vect{x}') = V^{(0)} (\vect{x}, \vect{x}') + V^{(1)}
   (\vect{x}, \vect{x}') + V^{(2)} (\vect{x}, \vect{x}') + \cdots
   . \label{eq:V_power}
\end{equation}
Here $V^{(0)} (\vect{x}, \vect{x}')$ is the self-consistent potential
of the reference crystal, which has the same periodicity as the ionic
potential (\ref{eq:V_ion_0}):
\begin{equation}
   V^{(0)}(\vect{x}, \vect{x}') = V^{(0)}(\vect{x} + \vect{R},
   \vect{x}' + \vect{R}) . \label{eq:V0}
\end{equation}
The linear response to the heterostructure perturbation has the form
\begin{equation}
   V^{(1)}(\vect{x}, \vect{x}') = \sideset{}{'} \sum_{\alpha,\vect{R}}
   \theta^{\alpha}_{\vect{R}} \Delta v^{\alpha}_{\vect{R}} (\vect{x},
   \vect{x}') , \label{eq:V1}
\end{equation}
while the quadratic response is
\begin{equation}
   V^{(2)}(\vect{x}, \vect{x}') = \sideset{}{'} \sum_{\alpha,\vect{R}}
   \sideset{}{'} \sum_{\alpha', \vect{R}'} \theta^{\alpha}_{\vect{R}}
   \theta^{\alpha'}_{\vect{R}'} \Delta v^{\alpha \alpha'}_{\vect{R}
   \vect{R}'} (\vect{x}, \vect{x}') .  \label{eq:V2}
\end{equation}
Here the expansion coefficients in the power series are defined by
\begin{equation} \label{eq:DVxx}
\begin{split}
   \Delta v^{\alpha}_{\vect{R}} (\vect{x}, \vect{x}') & =
   \frac{\partial V (\vect{x}, \vect{x}')}{\partial
   \theta^{\alpha}_{\vect{R}}} , \\ \Delta v^{\alpha
   \alpha'}_{\vect{R} \vect{R}'} (\vect{x}, \vect{x}') & = \frac12
   \frac{\partial^2 V (\vect{x}, \vect{x}')}{\partial
   \theta^{\alpha}_{\vect{R}} \; \partial
   \theta^{\alpha'}_{\vect{R}'}} ,
\end{split}
\end{equation}
where the derivatives are evaluated with respect to the reference
crystal.  These derivatives can be evaluated numerically by applying
perturbations (\ref{eq:Delta_V_ion}) with small values of
$\theta^{\alpha}_{\vect{R}}$.

The leading source of error in quadratic response theory lies in the
neglected cubic response term $V^{(3)}$.  This error can be reduced
somewhat by a suitable choice of reference potential.  For example, in
a GaAs/AlAs heterostructure, choosing
$\text{Al}_{0.5}\text{Ga}_{0.5}\text{As}$ as a reference crystal
instead of GaAs would reduce the cubic error in AlAs by a factor of 8.
However, it would also create a comparable error in GaAs.  Thus, the
cubic error in the interface Hamiltonian for an
$\text{Al}_{0.5}\text{Ga}_{0.5}\text{As}$ reference potential would be
about $\frac14$ that of a GaAs or AlAs reference potential.  This is
certainly an improvement, but since the quadratic response is already
quite small (see Sec.\ \ref{sec:est_quad}), it is unlikely to make
much practical difference.

In the momentum representation, the Dyson equation (\ref{eq:schr_x})
has the form
\begin{equation}
   \frac12 k^2 \psi(\vect{k}) + \sum_{\vect{k}'} V(\vect{k},
   \vect{k}') \psi(\vect{k}') = E \psi(\vect{k}) . \label{eq:schr_q}
\end{equation}
Making use of the translation symmetry of the reference crystal, one
can write the Fourier transforms of the potentials (\ref{eq:DVxx}) as
\begin{equation}
   \begin{split}
   \Delta v^{\alpha}_{\vect{R}} (\vect{k}, \vect{k}') &
   \equiv e^{-i (\vect{k} - \vect{k}') \cdot \vect{R}_{\alpha}}
   \Delta v^{\alpha} (\vect{k}, \vect{k}') , \\
   \Delta v^{\alpha \alpha'}_{\vect{R}
   \vect{R}'} (\vect{k}, \vect{k}') & \equiv e^{-i (\vect{k} -
   \vect{k}') \cdot \vect{R}_{\alpha}} \Delta v^{\alpha
   \alpha' \vect{R}'' } (\vect{k}, \vect{k}') ,
   \end{split} \label{eq:Dvk}
\end{equation}
in which $\vect{R}'' = \vect{R}' - \vect{R}$, and the coordinate
origin is $\vect{R}_{\alpha}$ for the modified functions on the
right-hand side.  The Fourier transform of the linear response
(\ref{eq:V1}) can therefore be written as
\begin{equation}
   V^{(1)}(\vect{k}, \vect{k}') = N \sideset{}{'} \sum_{\alpha}
   \theta^{\alpha} (\vect{k} - \vect{k}') \Delta v^{\alpha} (\vect{k},
   \vect{k}') .  \label{eq:V1kk}
\end{equation}
Here $N = \Omega / \Omega_0$ is the number of unit cells (of volume
$\Omega_0$) in the reference crystal (of volume $\Omega$),
$\theta^{\alpha} (\vect{k})$ is the Fourier transform of
$\theta^{\alpha}_{\vect{R}}$:
\begin{subequations}
\begin{align}
   \theta^{\alpha}_{\vect{R}} & = \sum_{\vect{k} \in \Omega_0^*}
   \theta^{\alpha} (\vect{k}) e^{i \vect{k} \cdot \vect{R}_{\alpha}} ,
   \label{eq:four_theta_a} \\ \theta^{\alpha} (\vect{k}) & =
   \frac{1}{N} \sum_{\vect{R} \in \Omega} \theta^{\alpha}_{\vect{R}}
   e^{-i \vect{k} \cdot \vect{R}_{\alpha}} , \label{eq:four_theta_b}
\end{align}
\end{subequations}
and $\Omega_0^* = (2\pi)^3 / \Omega_0$ is the volume of a unit cell in
the reciprocal lattice.  Note from (\ref{eq:four_theta_b}) that
$\theta^{\alpha} (\vect{k})$ is quasi-periodic:
\begin{equation}
   \theta^{\alpha} (\vect{k} + \vect{G}) = \theta^{\alpha} (\vect{k})
   e^{-i \vect{G} \cdot \bm{\tau}_{\alpha}} , \label{eq:quasi_per}
\end{equation}
where $\vect{G}$ is a reciprocal lattice vector of the reference
crystal.

In a similar fashion, inserting (\ref{eq:Dvk}) into the Fourier
transform of (\ref{eq:V2}) and relabeling $\vect{R}'$ gives
\begin{equation}
   V^{(2)}(\vect{k}, \vect{k}') = N \sideset{}{'} \sum_{\alpha,
   \alpha', \vect{R}'} \theta^{\alpha \alpha' \vect{R}' } (\vect{k} -
   \vect{k}') \Delta v^{\alpha \alpha' \vect{R}' } (\vect{k},
   \vect{k}') , \label{eq:V2kk}
\end{equation}
in which $\theta^{\alpha \alpha' \vect{R}' } (\vect{k})$ is the
Fourier transform of the pair distribution function $\theta^{\alpha
\alpha' \vect{R}'} (\vect{R}) \equiv \theta^{\alpha}_{\vect{R}}
\theta^{\alpha' }_{\vect{R} + \vect{R}'}$.

\subsection{Estimation of magnitude}

\label{sec:est_quad}

A crude estimate of the relative magnitudes of the linear and
quadratic response can be obtained from a simple nonlinear
Thomas-Fermi model, \cite{HoKo64} with the result \cite{Fore_unpub}
\begin{equation}
   V^{(1)} \sim \Delta \bar{V}, \qquad \frac{V^{(2)}}{V^{(1)}} \sim
   \frac{\Delta \bar{V}}{4 \epsilon_{\text{F}}} , \label{eq:qrb}
\end{equation}
where $\Delta \bar{V}$ is the typical band offset defined in Sec.\
\ref{sec:error}, and $\epsilon_{\text{F}}$ is the Fermi energy.  
Now the difference in screened pseudopotentials between typical III-V
semiconductors is roughly $\Delta \bar{V} \sim 0.02$--0.05 Ry,
\cite{CoHe70} which yields the estimate $|V^{(2)} / V^{(1)} | \sim
0.01$.  This suggests that the quadratic response is indeed very
small.

This estimate is supported by the LDA calculations of Wang and Zunger
\cite{WaZu97} for GaAs/AlAs heterostructures, in which they found (see
Table I \cite{WaZu97}) that the interface band-mixing terms arising
from the linear response were on average about 1000 times larger than
those arising from the quadratic response (with the ratio between the
smallest linear and largest quadratic terms being about 100, in
agreement with the estimate obtained above).  However, the LDA
calculations for GaAs/AlAs and In$_{0.53}$Ga$_{0.47}$As/InP presented
in Refs.\ \onlinecite{BaReBaPe89} and \onlinecite{PBBR90} indicate
that the quadratic density response is only about 10 times smaller
than the linear density response (the linear and quadratic potentials
were not given in these papers).

Nevertheless, this is still a sufficiently large ratio to establish
the validity of the quadratic approximation used here.  For the
purposes of the perturbation scheme of Sec.\ \ref{sec:error}, the
factor $4 \epsilon_{\text{F}}$ in the denominator of Eq.\
(\ref{eq:qrb}) will be treated formally as of order
$\bar{E}_{\text{g}}$, so that $\Delta V^{(2)}$ is considered to be of
the same order [$(\Delta \bar{V})^2 / \bar{E}_{\text{g}} \sim (\bar{k}
a)^2 \Delta \bar{V}$] as the smallest terms retained in Sec.\
\ref{sec:error}.

\subsection{Functional form}

As shown in Ref.\ \onlinecite{Fore05a}, the Coulomb interaction gives
rise to singularities in the linear and quadratic potentials $\Delta v
(\vect{k}, \vect{k}')$ when $\vect{k} - \vect{k}'$ is equal to a
reciprocal lattice vector.  The explicit form of the linear potential
is \cite{Fore05a}
\begin{multline}
   \Delta v^{\alpha}_{\vect{R}} (\vect{k} + \vect{G},
   \vect{k}' + \vect{G}') = w^{\alpha}_{\vect{R}} (\vect{k},
   \vect{k}'; \vect{G}, \vect{G}') \\ + \Lambda_{\vect{G} \vect{G}'}
   (\vect{k}, \vect{k}') \varphi^{\alpha}_{\vect{R}}
   (\vect{k} - \vect{k}') ,
\end{multline}
where the potential $w^{\alpha}_{\vect{R}} (\vect{k}, \vect{k}';
\vect{G}, \vect{G}')$ is an analytic function of $\vect{k}$ and
$\vect{k}'$ for wave vectors inside the first Brillouin zone of the
reference crystal.  For small values of $\vect{k}$ and $\vect{k}'$ it
can be represented as a Taylor series, which is the basis for the
effective-mass theory developed in the following section.  The
effective vertex function $\Lambda_{\vect{G} \vect{G}'} (\vect{k},
\vect{k}')$ is also an analytic function of $\vect{k}$ and
$\vect{k}'$.

The singular contributions come from the screened potential $\varphi$,
which is a spin scalar of the form \cite{Fore05a,note:new_phi}
\begin{equation}
   \varphi^{\alpha}_{\vect{R}} (\vect{q}) = \frac{v_c (\vect{q})
   n^{\alpha}_{\vect{R}} (\vect{q})}{\epsilon (\vect{q})} ,
   \label{eq:phi1}
\end{equation}
where $\epsilon (\vect{q})$ is the static electronic dielectric
function of the reference crystal and
\begin{equation}
   v_c(\vect{q}) = \begin{cases}
      4 \pi / q^2  & \text{if $q \ne 0$,} \\
      0 & \text{if $q = 0$.}
   \end{cases}
   \label{eq:vc}
\end{equation}
The function $n^{\alpha}_{\vect{R}} (\vect{q})$ is an effective
electron density containing partial contributions from the bare ionic
pseudocharge and the screening charge.  This is an analytic function
of $\vect{q}$ (for small $q$) with the symmetry of site
$\vect{R}_{\alpha}$ in the reference crystal.  For isovalent
substitutions in zinc-blende crystals, the leading nonanalytic terms
in Eq.\ (\ref{eq:phi1}) are octopole and hexadecapole potentials
proportional to $q_x q_y q_z / q^2$ and $(q_x^4 + q_y^4 + q_z^4) /
q^2$. \cite{Fore05a}

The quadratic potential has a similar form:
\begin{multline}
   \Delta v^{\alpha \alpha'}_{\vect{R} \vect{R}'} (\vect{k} +
   \vect{G}, \vect{k}' + \vect{G}') = w^{\alpha \alpha'}_{\vect{R}
   \vect{R}'} (\vect{k}, \vect{k}'; \vect{G}, \vect{G}') \\ +
   \Lambda_{\vect{G} \vect{G}'} (\vect{k}, \vect{k}') \varphi^{\alpha
   \alpha' }_{\vect{R} \vect{R}'} (\vect{k} - \vect{k}') ,
\end{multline}
in which $\Lambda$ is the same vertex function as above.  In this
case, however, the potential $w^{\alpha \alpha'}_{\vect{R} \vect{R}'}
(\vect{k}, \vect{k}'; \vect{G}, \vect{G}')$ is not an analytic
function of $\vect{k}$ and $\vect{k}'$.  Nevertheless, to within the
accuracy required here (namely, zeroth order in $\vect{k}$ and
$\vect{k}'$), the nonanalytic part can be neglected for neutral
perturbations.  Note that the analytic part of the quadratic potential
is analytic over a smaller region in $\vect{k}$ space than the linear
potential.  Whereas $w^{\alpha }_{\vect{R}} (\vect{k}, \vect{k}';
\vect{G}, \vect{G}')$ was analytic for $\vect{k}$ and $\vect{k}'$
inside the first Brillouin zone, the analytic part of $w^{\alpha
\alpha'}_{\vect{R} \vect{R}'} (\vect{k}, \vect{k}'; \vect{G},
\vect{G}')$ is analytic only over the inner ``half'' of the Brillouin
zone (i.e., over the Brillouin zone of a crystal whose lattice
constants are double those of the reference crystal).

The quadratic screened potential is defined by
\begin{equation}
   \varphi^{\alpha \alpha'}_{\vect{R} \vect{R}'} (\vect{q}) =
   \frac{v_c (\vect{q}) n^{\alpha \alpha'}_{\vect{R} \vect{R}'}
   (\vect{q})}{\epsilon (\vect{q})} , \label{eq:phi2}
\end{equation}
in which the effective density $n^{\alpha \alpha'}_{\vect{R}
\vect{R}'} (\vect{q})$ is likewise not analytic, but can be
approximated as such.  The leading terms here are dipole and
quadrupole potentials (since the monopole term vanishes for an
insulator at zero temperature).  In zinc-blende crystals, only the
interface dipole term is non-negligible under the approximation scheme
used in this paper.

For use in Eqs.\ (\ref{eq:V1kk}) and (\ref{eq:V2kk}), one requires
also expressions for the modified linear and quadratic potentials
defined in Eq.\ (\ref{eq:Dvk}).  These are given by
\begin{subequations} \label{eq:Dv_a}
\begin{multline}
   \Delta v^{\alpha} (\vect{k} + \vect{G}, \vect{k}' + \vect{G}') =
   w^{\alpha}_{\vect{G} \vect{G}'} (\vect{k}, \vect{k}') \\ + e^{i
   (\vect{G} - \vect{G}') \cdot \bm{\tau}_{\alpha}} \Lambda_{\vect{G}
   \vect{G}'} (\vect{k}, \vect{k}') \varphi^{\alpha} (\vect{k} -
   \vect{k}') , \label{eq:Dv1_a}
\end{multline}
\begin{multline}
   \Delta v^{\alpha \alpha' \vect{R}' } (\vect{k} + \vect{G},
   \vect{k}' + \vect{G}') = w^{\alpha \alpha' \vect{R}' }_{\vect{G}
   \vect{G}'} (\vect{k}, \vect{k}') \\ + e^{i (\vect{G} - \vect{G}')
   \cdot \bm{\tau}_{\alpha}} \Lambda_{\vect{G} \vect{G}'} (\vect{k},
   \vect{k}') \varphi^{\alpha \alpha' \vect{R}' } (\vect{k} -
   \vect{k}') , \label{eq:Dv2_a}
\end{multline}
\end{subequations}
in which the terms on the right-hand side are defined by expressions
similar to (\ref{eq:Dvk}).

\section{Envelope-function equations}

\label{sec:envelope}

In this section, the Dyson equation (\ref{eq:schr_q}) for a
heterostructure is written in the Luttinger-Kohn
rep\-re\-sen\-ta\-tion, \cite{LuKo55} in which the basis functions are
defined to be plane waves multiplied by the zone-center Bloch
functions $U_n$ of the periodic reference
crystal. \cite{Leib75,Burt88a} Here the set $\{ U_n \}$ is chosen to
be a complete set of solutions to the equation [cf.\ Eq.\
(\ref{eq:schr_x})]
\begin{multline}
   - \tfrac12 \nabla^2 U_n(\vect{x}, \omega) + \int V^{(\text{r},0)}
   (\vect{x}, \vect{x}', \omega) U_n(\vect{x}', \omega) d^3 \! x' \\ =
   E_n(\omega) U_n(\vect{x}, \omega) \label{eq:Un}
\end{multline}
that satisfy the periodic boundary conditions
\begin{equation}
   U_n(\vect{x}, \omega) = U_n(\vect{x} + \vect{R}, \omega)
   , \label{eq:pbc}
\end{equation}
in which $V^{(\text{r},0)}$ is the hermitian part of the reference
potential $V^{(0)}$.  For a general operator $A$, the ``real''
(hermitian) and ``imaginary'' (antihermitian) parts are defined here
as
\begin{equation}
   A^{\text{(r)}} = \frac12 (A^{\dag} + A) , \qquad
   A^{\text{(i)}} = \frac{i}{2} (A^{\dag} - A) .
\end{equation}

The solutions to Eq.\ (\ref{eq:Un}) form a complete orthonormal set of
periodic functions \cite{LuKo55} for any value of $\omega$.  In the
Fourier series representation
\begin{equation}
   U_n(\vect{x}, \omega) = \sum_{\vect{G}} U_{n\vect{G}} (\omega) e^{i
   \vect{G} \cdot \vect{x}} ,
\end{equation}
the orthogonality and completeness relations are
\begin{subequations} \label{eq:UnG_basis}
\begin{align}
   \sum_{\vect{G}} U_{n\vect{G}}^{\dag} (\omega) U_{n'\vect{G}}
   (\omega) & = \delta_{nn'} , \\ \sum_n U_{n\vect{G}} (\omega)
   U_{n\vect{G}'}^{\dag} (\omega) & = \openone \,
   \delta_{\vect{G}\vect{G}'} ,
\end{align}
\end{subequations}
in which $U_{n\vect{G}}^{\dag}$ denotes the hermitian conjugate of the
spinor $U_{n\vect{G}}$.

Note that the Bravais lattice chosen for the periodic boundary
conditions in Eq.\ (\ref{eq:pbc}) need not be the same as that in
(\ref{eq:V0}); for certain applications it may be preferable to impose
periodicity with respect to some (mathematically defined) supercell
instead.  For example, in treating intervalley $\Gamma$--$X$ coupling
in semiconductors with the zinc-blende structure, it is convenient
\cite{Burt92,Fore98b} to choose a nonprimitive simple cubic unit cell
of volume $\Omega_0 = a^3$ (where $a$ is the conventional cubic
lattice constant), which encompasses four primitive fcc
cells. \cite{note:nonorthogonal} This folds the $X$ valleys onto the
Brillouin zone center of the supercell, thereby permitting intervalley
$\Gamma$--$X$ mixing to be described in the same notation as that for
ordinary $\Gamma$ states (although the $\Gamma$ and $X$ states are of
course not coupled by the $\vect{k} \cdot \vect{p}$ interaction).

The wave function in the Luttinger-Kohn representation (or envelope
function) $F_n (\vect{k})$ is defined as
\begin{subequations} \label{eq:Fnk}
\begin{align}
   & F_n(\vect{k}) = \sum_{\vect{G}} U_{n\vect{G}}^{\dag}
   \psi(\vect{k} + \vect{G}) , \\ & \psi(\vect{k} + \vect{G})
   = \sum_{n} F_n(\vect{k}) U_{n\vect{G}} . \label{eq:psi_kG}
\end{align}
\end{subequations}
This definition is valid for any value of $\vect{k}$, but since only
those values from one unit cell $\Omega_0^*$ are needed to determine
$\psi(\vect{x})$, it is convenient to set $F_n(\vect{k}) \equiv 0$
when $\vect{k} \notin \Omega_0^*$. \cite{Burt88a,note:umklapp} The
Fourier transform of (\ref{eq:psi_kG}) is then the usual exact
envelope-function expansion \cite{LuKo55,Burt88a}
\begin{equation}
   \psi(\vect{x}) = \sum_n F_n(\vect{x}) U_n(\vect{x}) .
\end{equation}

The Dyson equation in the Luttinger-Kohn representation is given by
Eqs.\ (\ref{eq:schr_q}), (\ref{eq:UnG_basis}), and (\ref{eq:Fnk}) as
\begin{multline}
   E_n F_n (\vect{k}) + \sum_{n'} L_{nn'} (\vect{k}) F_{n'}
   (\vect{k}) \\ + \sum_{n'} \sum_{\vect{k}' \in \Omega_0^*}
   \Delta V_{nn'}(\vect{k}, \vect{k}') F_{n'} (\vect{k}') = E
   F_n (\vect{k}) . \label{eq:ef2}
\end{multline}
Here $\Delta V = V - V^{(0)}$ is the perturbation due to the
heterostructure, the matrix elements of which are
\begin{equation}
   \Delta V_{nn'} (\vect{k}, \vect{k}') = \sum_{\vect{G},\vect{G}'}
   U_{n\vect{G}}^{\dag} \Delta V(\vect{k} + \vect{G}, \vect{k}' +
   \vect{G}') U_{n'\vect{G}'} . \label{eq:Vnn}
\end{equation}
The term $L_{nn'} (\vect{k})$ groups together all contributions from
the bulk reference crystal Hamiltonian except $E_n$; i.e.,
\begin{equation}
   L_{nn'} (\vect{k}) = V^{(0)}_{nn'} (\vect{k}) -
   V^{(\text{r},0)}_{nn'} (\vect{0}) + \vect{k} \cdot \vect{p}_{nn'} +
   \tfrac12 k^2 \delta_{nn'} , \label{eq:Snnk}
\end{equation}
in which $V^{(0)}_{nn'} (\vect{k}) = V^{(0)}_{nn'} (\vect{k},
\vect{k})$ is the potential energy of the reference crystal and
\begin{equation}
   \vect{p}_{nn'} = \sum_{\vect{G}} \vect{G} U_{n\vect{G}}^{\dag}
   U_{n'\vect{G}} \label{eq:pnn}
\end{equation}
is the momentum matrix of the reference crystal.

Within quadratic response theory, the perturbation $\Delta V_{nn'}
(\vect{k}, \vect{k}')$ is obtained by substituting Eqs.\
(\ref{eq:V1kk}), (\ref{eq:V2kk}), and (\ref{eq:Dv_a}) into Eq.\
(\ref{eq:Vnn}).  The result is
\begin{equation}
   \Delta V_{nn'} (\vect{k}, \vect{k}') = W_{nn'} (\vect{k},
   \vect{k}') + \Lambda_{nn'} (\vect{k}, \vect{k}') \varphi (\vect{k}
   - \vect{k}') , \label{eq:DV12}
\end{equation}
where the vertex function
\begin{equation}
   \Lambda_{nn'} (\vect{k}, \vect{k}') = \sum_{\vect{G}, \vect{G}'}
   U_{n\vect{G}}^{\dag} \Lambda_{\vect{G} \vect{G}'} (\vect{k},
   \vect{k}') U_{n'\vect{G}'}
\end{equation}
is again an analytic function of $\vect{k}$ and $\vect{k}'$.  The $W$
term is defined by $W = W^{(1)} + W^{(2)}$, where
\begin{gather}
   W^{(1)}_{nn'} (\vect{k}, \vect{k}') = \sideset{}{'} \sum_{\alpha}
   \theta^{\alpha} (\vect{k} - \vect{k}') W^{\alpha}_{nn'} (\vect{k},
   \vect{k}') , \label{eq:W1} \\ W^{\alpha}_{nn'} (\vect{k},
   \vect{k}') = N \sum_{\vect{G}, \vect{G}'} U_{n\vect{G}}^{\dag}
   w^{\alpha}_{\vect{G} \vect{G}'} (\vect{k}, \vect{k}')
   U_{n'\vect{G}'} e^{i (\vect{G}' - \vect{G}) \cdot
   \bm{\tau}_{\alpha}} . \label{eq:W1a}
\end{gather}
The screened potential $\varphi$ is defined by
\begin{equation}
   \varphi(\vect{q}) = \frac{v_c (\vect{q}) n (\vect{q})}{\epsilon
   (\vect{q})} , \label{eq:phi_q}
\end{equation}
where $n = n^{(1)} + n^{(2)}$ is an effective electron density for the
heterostructure perturbation:
\begin{equation}
   n^{(1)} (\vect{q}) = N \sideset{}{'} \sum_{\alpha} \theta^{\alpha}
   (\vect{q}) n^{\alpha} (\vect{q}) . \label{eq:n1_q}
\end{equation}
The quadratic contributions $W^{(2)}$ and $n^{(2)}$ are given by
obvious generalizations [see Eq.\ (\ref{eq:Dv_a})] of the above
expressions.  The physical interpretation of these results is
considered below.

\section{Power series expansions}

\label{sec:power_series}

In this section, power series expansions are used to obtain
approximate expressions for the Hamiltonian matrix elements in the
envelope-function equations (\ref{eq:ef2}).  This approximation is
justified by the existence
\cite{Burt92,Burt94c,Xia89,WaFrZu97,WaZu97,WaZu99} of slowly varying
envelope functions $F_n (\vect{x})$, for which $F_n (\vect{k})$ is
negligible unless $\vect{k}$ is small.  This expansion provides a
starting point for the development of an approximate effective-mass
theory, and also assists in the physical interpretation of the various
terms in the Hamiltonian.

\subsection{Reference crystal Hamiltonian}

The leading terms in the bulk Hamiltonian $L_{nn'}(\vect{k})$ of Eq.\
(\ref{eq:Snnk}) are derived from a Taylor series expansion of
$V^{(0)}_{nn'}(\vect{k})$, and are given through terms of the fourth
order in $k$ by
\begin{multline}
   L_{nn'}(\vect{k}) = k_{\lambda} \pi^{\lambda}_{nn'} + k_{\lambda}
   k_{\mu} \tilde{D}^{\lambda\mu}_{nn'} + k_{\lambda} k_{\mu}
   k_{\kappa} \tilde{C}^{\lambda\mu\kappa}_{nn'} \\ + k_{\lambda}
   k_{\mu} k_{\kappa} k_{\nu} \tilde{Q}^{\lambda\mu\kappa\nu}_{nn'} .
   \label{eq:Snn}
\end{multline}
Here a summation with respect to the Cartesian indices $\lambda, \mu,
\kappa, \text{and } \nu$ is implicit, and the coefficients are
\begin{equation} \label{eq:pi_D_C_Q}
\begin{split}
   \pi_{nn'}^{\lambda} & = p_{nn'}^{\lambda} + \left( \frac{\partial
   V^{(\text{r},0)}_{nn'} (\vect{k})}{\partial k_{\lambda}}
   \right)_{\vect{k} = \vect{0}} , \\
   \tilde{D}_{nn'}^{\lambda\mu} & = \frac12 \left( \delta_{\lambda\mu}
   \delta_{nn'} + \frac{\partial^2 V^{(\text{r},0)}_{nn'}
   (\vect{k})}{\partial k_{\lambda} \partial k_{\mu}}
   \right)_{\vect{k} = \vect{0}} , \\
   \tilde{C}_{nn'}^{\lambda\mu\kappa} & = \frac{1}{3!}  \left(
   \frac{\partial^3 V^{(\text{r},0)}_{nn'} (\vect{k})}{\partial
   k_{\lambda} \partial k_{\mu} \partial k_{\kappa}} \right)_{\vect{k}
   = \vect{0}} , \\
   \tilde{Q}_{nn'}^{\lambda\mu\kappa\nu} & = \frac{1}{4!}  \left(
   \frac{\partial^4 V^{(\text{r},0)}_{nn'} (\vect{k})}{\partial
   k_{\lambda} \partial k_{\mu} \partial k_{\kappa} \partial k_{\nu}}
   \right)_{\vect{k} = \vect{0}} .
\end{split}
\end{equation}
The various derivatives of $V^{(\text{r},0)}_{nn'} (\vect{k})$ account
for contributions from the nonlocal part of the potential energy to
the dispersion relation of the reference crystal.  The quantity
$\pi_{nn'}^{\lambda}$ is the kinetic momentum matrix of the reference
crystal, whereas the other terms give partial contributions (see Sec.\
\ref{sec:perturbation} for the remaining contributions) to the
effective-mass tensor and the cubic and quartic dispersion terms of
the reference crystal.

In principle, $L_{nn'}(\vect{k})$ should include contributions from
the antihermitian part of the self-energy $\Sigma^{\text{(i,0)}}_{nn'}
(\vect{k}, \omega) = V^{\text{(i,0)}}_{nn'} (\vect{k}, \omega)$.
However, in Appendix \ref{app:imag_sigma} it is shown that, for
energies $\omega$ near the band gap of the reference crystal, the
contributions from $\Sigma^{\text{(i)}}$ are much smaller than the
smallest terms retained in the present approximation scheme.  Such
contributions were therefore neglected in the above expressions, and
are likewise neglected in subsequent analysis of the heterostructure
perturbation.  However, in any calculation where it is desired to
include the effects of a finite quasiparticle lifetime, the dominant
terms may be restored to leading order by replacing $E_n$ with
\begin{equation}
   E_n(\omega) \rightarrow E_n(\omega) + i
   \Sigma^{(\text{i},0)}_{nn} (\vect{k} = \vect{0}, \omega) ,
   \label{eq:sig_nn}
\end{equation}
and then retaining the imaginary part only to first order in
perturbation theory in all subsequent analysis.

\subsection{Linear heterostructure potential}

A similar Taylor series expansion technique is useful for the terms
$W_{nn'} (\vect{k}, \vect{k}')$ and $\Lambda_{nn'} (\vect{k},
\vect{k}')$ in the heterostructure perturbation (\ref{eq:DV12}).  This
subsection begins by considering the simple special case in which the
screened atomic pseudopotentials $w^{\alpha}_{\vect{G}
\vect{G}'} (\vect{k}, \vect{k}')$ in Eqs.\ (\ref{eq:Dv1_a}) and
(\ref{eq:W1a}) have the form of a local potential; i.e.,
\begin{equation}
   w^{\alpha}_{\vect{G} \vect{G}'} (\vect{k}, \vect{k}') = w^{\alpha}
   (\vect{k} - \vect{k}' + \vect{G} - \vect{G}') . \label{eq:w_local}
\end{equation}
Such would be the case, for example, in an LDA calculation based on
local ionic pseudopotentials.  In this case, $W_{nn'}^{(1)} (\vect{k},
\vect{k}')$ is also a local potential of the form $W_{nn'}^{(1)}
(\vect{k}, \vect{k}') = W_{nn'}^{(1)} (\vect{k} - \vect{k}')$.  This
simplification makes it easier to grasp the physical significance of
the power series expansion, and also facilitates a comparison between
the present theory and earlier envelope-function models based on local
empirical pseudopotentials.

\subsubsection{Local analytic terms}

The approximation technique used here for $W^{(1)}_{nn'} (\vect{k} -
\vect{k}')$ is to expand the term $W^{\alpha}_{nn'} (\vect{k} -
\vect{k}')$ on the right-hand side of Eq.\ (\ref{eq:W1}) in a Taylor
series in $\vect{q} \equiv \vect{k} - \vect{k}'$:
\begin{equation}
   W^{(1)}_{nn'} (\vect{q}) = \sideset{}{'} \sum_{\alpha}
   \theta^{\alpha} (\vect{q}) [W^{\alpha}_{nn'} + i q_{\lambda}
   (Z^{\lambda}_{nn'})^{\alpha} - q_{\lambda} q_{\mu}
   (Y^{\lambda\mu}_{nn'})^{\alpha}] .  \label{eq:W1q}
\end{equation}
Here the terms $W^{\alpha}_{nn'}$, $(Z^{\lambda}_{nn'})^{\alpha}$, and
$(Y^{\lambda\mu}_{nn'})^{\alpha}$ are $\vect{q}$-in\-de\-pen\-dent
expansion coefficients.  The series (\ref{eq:W1q}) has been truncated
at the second order in $q$ because such terms are of order $(\bar{k}
a)^2 \Delta \bar{V}$, which are the smallest corrections permitted in
the present perturbation scheme.

The result (\ref{eq:W1q}) may be used directly in a $\vect{k}$-space
envelope-function calculation based on Eq.\ (\ref{eq:ef2}), in which
the $\vect{k}$ values are expressly limited to the unit cell
$\Omega_0^*$.  However, to obtain a local differential equation, one
must allow $\vect{k}$ to range over all possible values.  Since
$\theta^{\alpha} (\vect{k})$ is quasi-periodic [see Eq.\
(\ref{eq:quasi_per})], this local approximation will generate
large-$\vect{k}$ terms in the envelope functions unless a
$\vect{k}$-space cutoff is introduced.  If this is done, the Fourier
transform of (\ref{eq:W1q}) is the local potential
\begin{equation}
   W^{(1)}_{nn'} (\vect{x}) = \sideset{}{'} \sum_{\alpha} [W^{\alpha
   }_{nn'} + (Z^{\lambda}_{nn'})^{\alpha} \partial_{\lambda} +
   (Y^{\lambda\mu}_{nn'})^{\alpha} \partial_{\lambda} \partial_{\mu}]
   \theta^{\alpha} (\vect{x}) , \label{eq:W1x}
\end{equation}
in which $\partial_{\lambda} = \partial / \partial x_{\lambda}$ and
[cf.\ Eq.~(\ref{eq:four_theta_a})]
\begin{equation}
   \theta^{\alpha} (\vect{x}) = \sum_{\vect{k}} B (\vect{k})
   \theta^{\alpha} (\vect{k}) e^{i \vect{k} \cdot \vect{x}}
   , \label{eq:theta_x}
\end{equation}
where the cutoff function $B(\vect{k})$ is defined in Appendix
\ref{app:cutoff}.  From Eq.\ (\ref{eq:theta_constraint}), the
constraint
\begin{equation}
   \sum_{a} \theta^{aj} (\vect{x}) = 0 \label{eq:theta_x_constraint}
\end{equation}
is satisfied for any choice of $B(\vect{k})$.

The physical significance of the result (\ref{eq:W1x}) can be
appreciated by considering a specific example such as a (001)
GaAs/AlAs heterojunction.  In this case, as will be shown in Sec.\
\ref{sec:stepfunction}, the function $\theta^{\alpha} (\vect{x})$
depends only on the $z$ coordinate and behaves like a smooth step
function at the interface.  The spatial derivatives in Eq.\
(\ref{eq:W1x}) therefore generate finite-width $\delta$-like terms at
the interface, with the $Z$ term proportional to $\delta(z)$ and the
$Y$ term proportional to $\delta'(z)$.  Hence, Eq.\ (\ref{eq:W1x})
provides a first example of the interface band-mixing terms alluded to
previously in Eqs.\ (\ref{eq:d0}) and (\ref{eq:d1}).

The physical origin of these terms can be understood by going back one
step further in the derivation.  From Eq.\ (\ref{eq:W1a}), it is clear
that the Taylor series expansion of $W^{\alpha}_{nn'} (\vect{q})$ in
Eq.\ (\ref{eq:W1q}) is equivalent to a Taylor series expansion of the
screened atomic pseudopotential (\ref{eq:w_local}) with respect to
$\vect{q} = \vect{k} - \vect{k}'$.  Hence, the physical origin of the
linear and quadratic (in $q$) terms in Eq.\ (\ref{eq:W1q}) is simply
the finite slope and curvature of the screened atomic pseudopotentials
in momentum space.

This demonstrates that the Hamiltonian of a heterostructure depends
not just on the values of the atomic pseudopotentials $w^{\alpha}
(\vect{k})$ at the reciprocal lattice vectors $\vect{G}$, but also at
a {\em range} of $\vect{k}$ values in a finite neighborhood of each
$\vect{G}$ (the size of the neighborhood depending on how rapidly
varying the envelope function is).  The necessity for an accurate
fitting of empirical pseudopotentials over a range of $\vect{k}$
values has been emphasized particularly in the work of M\"ader and
Zunger. \cite{MaZu94} The truncated expansion (\ref{eq:W1q}) shows
that the present perturbation scheme relies for its accuracy upon the
validity of a quadratic extrapolation of $w^{\alpha} (\vect{k})$ in
the neighborhood of each $\vect{G}$.  Inspection of the form of
typical atomic pseudopotentials \cite{CoHe70,CoCh89} shows this to be
a good approximation.

The definition (\ref{eq:W1q}) of the interface band-mixing parameters
$Z$ and $Y$ highlights a significant difference between the present
approach and previous derivations of envelope-function Hamiltonians.
In previous derivations, the heterostructure potential was chosen to
have the form
\begin{equation}
   V (\vect{x}) = \sum_{l} V^{l} (\vect{x}) \theta^{l} (\vect{x}) ,
   \label{eq:Vmodel}
\end{equation}
in which $V^{l} (\vect{x})$ is the periodic potential for the bulk
material $l$, and $\theta^{l} (\vect{x})$ is a form factor determining
the composition of the heterostructure.  Within this model, $V
(\vect{x})$ depends only upon the atomic pseudopotentials at the
reciprocal lattice vectors $\vect{G}$, with the magnitude of the
interface $\delta$-like terms determined by $\theta^{l} (\vect{x})$.
In the limit of slowly varying $\theta^{l} (\vect{x})$, the interface
terms vanish.

The present results show that this behavior is an unphysical artifact
of the model (\ref{eq:Vmodel}).  In the present theory, the strength
of the interface terms is the same (in the virtual crystal
approximation) for slowly graded and abrupt heterostructures, with the
rate of change of $\theta^{\alpha} (\vect{x})$ affecting
only the width of the interface terms.

\subsubsection{Nonlocal analytic terms}

The more general case of a nonlocal potential involves only a
straightforward extension of these results.  The matrix element
$W^{\alpha}_{nn'} (\vect{k}, \vect{k}')$ of Eq.\ (\ref{eq:W1a}) is
expanded to second order in $\vect{k}$ and $\vect{k}'$ as follows:
\begin{multline}
   W^{\alpha}_{nn'} (\vect{k}, \vect{k}') = W^{\alpha}_{nn'} +
   k_{\lambda} (\tilde{J}^{\lambda}_{nn'})^{\alpha} + k_{\lambda}'
   [(\tilde{J}^{\lambda}_{n'n})^{\alpha}]^{*} \\ + k_{\lambda} k_{\mu}
   (\tilde{M}^{\lambda\mu}_{nn'})^{\alpha} + k_{\lambda}' k_{\mu}'
   [(\tilde{M}^{\mu\lambda}_{n'n})^{\alpha}]^{*} \\ + k_{\lambda}
   k_{\mu}' (\tilde{R}^{\lambda\mu}_{nn'})^{\alpha}
   . \label{eq:W1_Taylor}
\end{multline}
Here the antihermitian part of the self-energy was neglected, as
discussed above Eq.\ (\ref{eq:sig_nn}) and in Appendix
\ref{app:imag_sigma}.  This approximation simplifies the expansion by
providing a relationship between (for example) the coefficients of
$k_{\lambda}$ and $k_{\lambda}'$.  The physical interpretation of the
various expansion coefficients is discussed below (in Sec.\
\ref{sec:perturbation}), after perturbation theory has been used to
eliminate the coupling to remote bands.

When this expansion is substituted into Eq.\ (\ref{eq:W1}) for
$W^{(1)}_{nn'} (\vect{k}, \vect{k}')$, the result can be written as
\begin{multline}
   W^{(1)}_{nn'} (\vect{k}, \vect{k}') = W^{(1)}_{nn'} (\vect{q}) +
   k_{\lambda} \tilde{J}^{\lambda}_{nn'} (\vect{q}) + k_{\lambda}'
   [\tilde{J}^{\lambda}_{n'n} (-\vect{q})]^{*} \\ + k_{\lambda}
   k_{\mu} \tilde{M}^{\lambda\mu}_{nn'} (\vect{q}) + k_{\lambda}'
   k_{\mu}' [\tilde{M}^{\mu\lambda}_{n'n} (-\vect{q})]^{*} \\ +
   k_{\lambda} k_{\mu}' \tilde{R}^{\lambda\mu}_{nn'} (\vect{q}) ,
   \label{eq:W1kk}
\end{multline}
in which $\vect{q} = \vect{k} - \vect{k}'$ as before, and the various
$\vect{q}$-de\-pen\-dent functions are defined by expressions of the
form
\begin{equation}
   \tilde{R}^{\lambda\mu}_{nn'} (\vect{q}) = \sideset{}{'}
   \sum_{\alpha} \theta^{\alpha} (\vect{q})
   (\tilde{R}^{\lambda\mu}_{nn'})^{\alpha} . \label{eq:Rq}
\end{equation}
These functions all have a step-function-like behavior in $\vect{x}$
space at a heterojunction (see Sec.\ \ref{sec:stepfunction}).  Since
$\theta^{\alpha}_{\vect{R}}$ is real, they also have the hermiticity
and symmetry properties
\begin{equation} \label{eq:Psym}
\begin{split}
   [ W^{(1)}_{nn'} (\vect{q})]^{*} & = W^{(1)}_{n'n} (-\vect{q}) , \\
   [ \tilde{R}^{\lambda\mu}_{nn'} (\vect{q}) ]^{*} & =
   \tilde{R}^{\mu\lambda}_{n'n} (-\vect{q}) , \\
   \tilde{M}^{\lambda\mu}_{nn'} (\vect{q}) & =
   \tilde{M}^{\mu\lambda}_{nn'} (\vect{q}) .
\end{split}
\end{equation}

\subsubsection{Nonanalytic terms}

The term $\Lambda_{nn'} (\vect{k}, \vect{k}') \varphi^{(1)}
(\vect{q})$ describing the nonanalytic contributions to Eq.\
(\ref{eq:DV12}) is handled in much the same way.  The vertex function
$\Lambda_{nn'} (\vect{k}, \vect{k}')$ is expanded in a Taylor series
of the form (\ref{eq:W1_Taylor}), while the linear electron density in
Eq.\ (\ref{eq:n1_q}) is expanded as
\begin{equation}
   n^{\alpha} (\vect{q}) = q_{\lambda} n^{\alpha}_{\lambda} +
   q_{\lambda} q_{\mu} n^{\alpha}_{\lambda\mu} + q_{\lambda} q_{\mu}
   q_{\kappa} n^{\alpha}_{\lambda\mu\kappa} + q_{\lambda} q_{\mu}
   q_{\kappa} q_{\nu} n^{\alpha}_{\lambda\mu\kappa\nu} ,
   \label{eq:n1_Taylor}
\end{equation}
which is valid for neutral perturbations.  In a heterovalent
zinc-blende heterostructure described by neutral perturbations,
\cite{Fore05a,note:heterovalent} the dipole and traceless quadrupole
terms $n^{\alpha}_{\lambda}$ and $n^{\alpha}_{\lambda\mu} - \frac13
n^{\alpha}_{\nu\nu} \delta_{\lambda\mu}$ are nonvanishing only at
interfaces.  For isovalent substitutions in zinc-blende,
$n^{\alpha}_{\lambda} = 0$ and $n^{\alpha}_{\lambda\mu} = \frac13
n^{\alpha}_{\nu\nu} \delta_{\lambda\mu}$ everywhere, so the
contribution from the latter term is analytic and can be absorbed into
the definition of the analytic potential (\ref{eq:W1_Taylor}).  The
remaining terms $n^{\alpha}_{\lambda\mu\kappa}$ and
$n^{\alpha}_{\lambda\mu\kappa\nu}$ describe octopole and hexadecapole
moments. \cite{Fore05a}

The inverse dielectric function $\epsilon^{-1} (\vect{q})$ in Eq.\
(\ref{eq:phi_q}) can also be expanded in a power series.  For
isovalent perturbations in zinc-blende this has no qualitative
significance, since the leading correction merely renormalizes the
hexadecapole term $n^{\alpha}_{\lambda\mu\kappa\nu}$.  \cite{Fore05a}
However, the wave vector dependence of $\epsilon^{-1} (\vect{q})$ does
generate a qualitatively new contribution from the interface dipole
term in Eq.\ (\ref{eq:n1_Taylor}); see Eq.\ (6.23) of Ref.\
\onlinecite{Fore05a} for the explicit form of this term.

With the expansion (\ref{eq:n1_Taylor}), the effective density
(\ref{eq:n1_q}) has the same form as that derived above for the local
potential in Eq.\ (\ref{eq:W1q}).  In coordinate space, it involves a
series of derivatives of $\theta^{\alpha} (\vect{x})$, similar to the
result shown in Eq.\ (\ref{eq:W1x}).  This is the same as the usual
multipole expansion of the macroscopic charge density in classical
electromagnetism. \cite{Jackson99_p253} In particular, note that for a
perturbation consisting of a single impurity atom, the function
$\theta^{\alpha} (\vect{x})$ is just the macroscopic average of a
Dirac $\delta$ function, in complete agreement with the expressions
given in Ref.\ \onlinecite{Jackson99_p253}.

Since the Taylor series expansion for $\Lambda_{nn'} (\vect{k},
\vect{k}')$ is identical in form to that for $W^{\alpha}_{nn'}
(\vect{k}, \vect{k}')$, the contribution from this term will not be
written out explicitly here.  All subsequent perturbation theory
analysis for the two terms is formally identical, except that the
vertex function is multiplied by an extra factor of
$\varphi(\vect{q})$.

\subsection{Quadratic heterostructure potential}

\subsubsection{Analytic terms}

\label{sec:quad_an}

Because the quadratic response is already of order $(\Delta \bar{V})^2
/ \bar{E}_{\text{g}}$, the quadratic version of Eq.\ (\ref{eq:W1}) can
be replaced by the zeroth-order approximation
\begin{align}
   W^{(2)}_{nn'} (\vect{k}, \vect{k}') & = \sideset{}{'}
   \sum_{\alpha,\alpha',\vect{R}'} \theta^{\alpha\alpha'\vect{R}'}
   (\vect{q}) W^{\alpha\alpha'\vect{R}'}_{nn'} \nonumber \\ & \equiv
   \tilde{W}^{(2)}_{nn'} (\vect{q}) , \label{eq:W2b}
\end{align}
in which 
\begin{equation}
   W^{\alpha\alpha'\vect{R}'}_{nn'} = \lim_{k, k' \rightarrow 0}
   W^{\alpha\alpha'\vect{R}'}_{nn'} (\vect{k}, \vect{k}') ,
\end{equation}
where the limit is well defined for neutral
per\-tur\-ba\-tions. \cite{Fore05a} This contributes a local
potential-energy term similar to that given by $W^{(1)}_{nn'}
(\vect{q})$ in Eq.\ (\ref{eq:W1kk}).  At a heterojunction, the
functional dependence in $\vect{x}$ space is similar to that of a
smooth (macroscopically averaged) step function, with possible
deviations in the vicinity of the junction.

Now since $W^{\alpha\alpha'\vect{R}'}_{nn'}$ depends only on the part
of $W^{\alpha\alpha'\vect{R}'}_{nn'} (\vect{k}, \vect{k}')$ that is
analytic in $\vect{k}$ and $\vect{k}'$, it is a short-range quantity
that is significant only when $\vect{R}' + \bm{\tau}_{\alpha'} -
\bm{\tau}_{\alpha}$ is comparable to the lattice constant $a$.
Therefore, at a heterojunction, replacing $\tilde{W}^{(2)}_{nn'}
(\vect{q})$ with an ideal step function would generate an error of
order $\bar{k} a$ in a term of order $(\bar{k} a)^2 \Delta \bar{V}$.
Hence, the error is of the same order as the local approximation
(\ref{eq:local_error}) and can be neglected.

Thus, within the present perturbation scheme there is no interface
contribution from $\tilde{W}^{(2)}_{nn'} (\vect{q})$.  This is an
important result, as it simplifies the analysis of interface effects
in later sections of this paper.

Within the context of an empirical pseudopotential model,
\cite{MaZu94} the nonlinear bulk term $\tilde{W}^{(2)}_{nn'}
(\vect{q})$ derived here can be represented as an environment
dependence of the screened empirical pseudopotential, in which (for
example) the pseudopotential for an As atom in GaAs is different from
that for an As atom in AlAs.  The importance of accounting for such
effects has been emphasized by M\"ader and Zunger. \cite{MaZu94}

\subsubsection{Nonanalytic terms}

The leading terms in the quadratic density are the dipole and
quadrupole terms
\begin{equation}
   n^{\alpha\alpha'\vect{R}'} (\vect{q}) = q_{\lambda}
   n^{\alpha\alpha'\vect{R}'}_{\lambda} + q_{\lambda} q_{\mu}
   n^{\alpha\alpha'\vect{R}'}_{\lambda\mu} , \label{eq:n2_Taylor}
\end{equation}
which again is valid for neutral perturbations.  (Here the constant
term vanishes even for charged perturbations as long as the system is
insulating, \cite{Fore05a} but in this case there is an additional
nonanalytic term of order $q^2$.\cite{Fore05a}) In zinc-blende
materials, the net contributions to $n^{(2)} (\vect{q})$ from
$n^{\alpha\alpha'\vect{R}'}_{\lambda}$ and the traceless part of
$n^{\alpha\alpha'\vect{R}'}_{\lambda\mu}$ both vanish in the bulk
regions of a heterostructure.  The interface part of
$n^{\alpha\alpha'\vect{R}'}_{\lambda\mu}$ is negligible under the
current approximation scheme, while the bulk part can be absorbed into
the definition of the analytic potential $\tilde{W}^{(2)}_{nn'}
(\vect{q})$.  Therefore, only the interface dipole term remains under
the current approximation scheme.

\subsection{Two-dimensional systems}

\label{sec:2d}

In a heterostructure (such as a quantum well) with two-dimensional
translation symmetry, the nonanalytic terms arising from $\varphi
(\vect{q})$ have a particularly simple form.  Let the dimensionless
coordinates $x_i$ and $k_j$ be defined by $\vect{x} = x_i \vect{a}_i$
and $\vect{k} = k_j \vect{b}_j$, where $\vect{a}_i$ and $\vect{b}_j$
are basis vectors for the direct and reciprocal lattices of the
reference crystal, with $\vect{a}_i \cdot \vect{b}_j = 2 \pi
\delta_{ij}$.  In these coordinates, the lattice sites are defined by
$x_i = R_i$ and $k_j = G_j$, where $R_i$ and $G_j$ are integers.  In
such a two-dimensional system, one can choose $\vect{a}_1$ and
$\vect{a}_2$ to lie parallel to the junction plane, so that the atomic
distribution function $\theta^{\alpha}_{\vect{R}}$ is independent of
$R_1$ and $R_2$.  The Fourier transform $\theta^{\alpha} (\vect{k})$
then has the form
\begin{equation}
   \theta^{\alpha} (\vect{k}) = \theta^{\alpha} (k_3)
   \sum_{\vect{G}_{\parallel}} \delta_{\vect{k}_{\parallel}
   \vect{G}_{\parallel}} e^{-i \vect{G}_{\parallel} \cdot
   \bm{\tau}_{\alpha}} , \label{eq:theta_2d}
\end{equation}
where $\vect{k}_{\parallel} = k_1 \vect{b}_1 + k_2 \vect{b}_2$.  For
small $\vect{k}_{\parallel}$, the $\vect{k}_{\parallel}$ dependence is
simply $\delta_{\vect{k}_{\parallel} \vect{0}}$.  The same conclusion
holds for the pair distribution function
$\theta^{\alpha\alpha'\vect{R}'} (\vect{k})$.

Therefore, the nonanalytic potential (\ref{eq:phi_q}) has the form
$\varphi (\vect{q}) = \varphi (q_3) \delta_{\vect{q}_{\parallel}
\vect{0}}$, which is independent of the direction of $\vect{q}$.  As a
result, all terms in $\varphi (\vect{q})$ except the monopole and
dipole terms reduce to analytic functions of $q_3$, which can be
absorbed into the definition of the analytic potential $W$.  In regard
to the monopole and dipole terms, this paper considers only neutral
perturbations in crystals with no bulk dipole moment.  Thus, the only
nonanalytic contributions are the $1/q_3$ terms generated by the
interface dipoles in Eqs.\ (\ref{eq:n1_Taylor}) and
(\ref{eq:n2_Taylor}).

These have the same $q_3$ dependence as the Fourier transform of a
step function, and merely add extra terms to the band offsets at a
heterojunction.  Hence, the interface dipole contributions can be
absorbed into the definition of $W^{(1)}_{nn'} (\vect{q})$ and
$\tilde{W}^{(2)}_{nn'} (\vect{q})$.  Note that in a no-common-atom
system such as InAs/GaSb, the contribution from the quadratic
interface dipole to the band offset has a different value for
GaAs-like and InSb-like interfaces. \cite{DanZun92}

In summary, for the material systems considered in this paper, the
nonanalytic potential $\varphi(\vect{q})$ does not contribute anything
qualitatively new in a heterostructure with two-dimensional
translation symmetry.  Only the interface dipole term is truly
nonanalytic, and that can be absorbed into the definition of the band
offsets (although this contribution does depend on the microscopic
structure of the interface\cite{DanZun92}).  Long-range potentials
arising from the direction dependence of $\varphi(\vect{q})$ appear
only in structures with lower translation symmetry, such as quantum
wires and dots.

\section{Elimination of interband coupling}

\label{sec:perturbation}

In this section, perturbation theory \cite{LuKo55,BirPik74,Wink03} is
used to derive a multi-band effective-mass Hamiltonian from the
in\-fi\-nite-dimen\-sion\-al matrix equations (\ref{eq:ef2}).  The
method is outlined briefly here; for further details see Refs.\
\onlinecite{BirPik74} and \onlinecite{Wink03}.

The zone-center states $n$ of the reference crystal are divided into a
class $A$ containing the states of interest, and a class $B$
containing all other states.  The total Hamiltonian is written as $H =
H_0 + H'$, where $H_0$ has matrix elements $(H_0)_{mm'} = E_m
\delta_{mm'}$, and $m = (n, \vect{k})$ is a composite index.  The
unperturbed Hamiltonian $H_0$ is assumed to be hermitian, but the
perturbation $H'$ need not be.  A similarity transformation $\bar{H} =
e^{-S} H e^S$ is used to eliminate the coupling between sets $A$ and
$B$ to any desired order in the perturbation $H'$.  (The
transformation is unitary if $H'$ is hermitian, as is approximately
the case here.)  This yields a finite-dimensional effective
Hamiltonian for states $m, m' \in A$, which is given explicitly (to
third order in $H'$) in Eq.\ (\ref{eq:bp138}) of Appendix
\ref{app:perturbation}.

For the case considered here, the matrix elements of $H_0$ and $H'$
are given by Eq.\ (\ref{eq:ef2}) as $\langle n \vect{k} | H_0 | n'
\vect{k}' \rangle = E_n \delta_{nn'} \delta_{\vect{k} \vect{k}'}$ and
\begin{multline}
   \langle n \vect{k} | H' | n' \vect{k}' \rangle = L_{nn'} (\vect{k})
   \delta_{\vect{k} \vect{k}'} + W^{(1)}_{nn'} (\vect{k}, \vect{k}') +
   \tilde{W}^{(2)}_{nn'} (\vect{q}) \\ + \Lambda_{nn'} (\vect{k},
   \vect{k}') \varphi (\vect{q}) ,
\end{multline}
in which $L_{nn'} (\vect{k})$ is defined in Eq.\ (\ref{eq:Snn}),
$W^{(1)}_{nn'} (\vect{k}, \vect{k}')$ in Eq.\ (\ref{eq:W1kk}), and
$\tilde{W}^{(2)}_{nn'} (\vect{q})$ in Eq.\ (\ref{eq:W2b}).  Upon
inserting these matrix elements into Eq.\ (\ref{eq:bp138}), one
obtains the effective-mass Hamiltonian (for $n, n' \in A$)
\begin{widetext}
\begin{multline}
   \langle n \vect{k} | \bar{H} | n' \vect{k}' \rangle = (E_n
   \delta_{nn'} + k_{\lambda} \pi^{\lambda}_{nn'} + k_{\lambda}
   k_{\mu} D^{\lambda\mu}_{nn'} + k_{\lambda} k_{\mu} k_{\kappa}
   C^{\lambda\mu\kappa}_{nn'} + k_{\lambda} k_{\mu} k_{\kappa} k_{\nu}
   Q^{\lambda\mu\kappa\nu}_{nn'} ) \delta_{\vect{kk}'} +
   W^{(1)}_{nn'} (\vect{q}) + W^{(2)}_{nn'} (\vect{q}) \\ +
   k_{\lambda} J^{\lambda}_{nn'} (\vect{q}) + k_{\lambda}'
   [J^{\lambda}_{n'n} (-\vect{q})]^{*} + k_{\lambda} k_{\mu}
   M^{\lambda\mu}_{nn'} (\vect{q}) + k_{\lambda}' k_{\mu}'
   [M^{\mu\lambda}_{n'n} (-\vect{q})]^{*} + k_{\lambda} k_{\mu}'
   R^{\lambda\mu}_{nn'} (\vect{q}) \\ + [ \Lambda_{nn'} + k_{\lambda}
   \hat{J}^{\lambda}_{nn'} + k_{\lambda}'
   (\hat{J}^{\lambda}_{n'n})^{*} + k_{\lambda} k_{\mu}
   \hat{M}^{\lambda\mu}_{nn'} + k_{\lambda}' k_{\mu}'
   (\hat{M}^{\mu\lambda}_{n'n})^{*} + k_{\lambda} k_{\mu}'
   \hat{R}^{\lambda\mu}_{nn'} ] \varphi(\vect{q}) . \label{eq:em1}
\end{multline}
\end{widetext}
In this expression, $\pi^{\lambda}_{nn'}$ is the kinetic momentum
matrix (\ref{eq:pi_D_C_Q}) of the reference crystal, $2
D^{\lambda\mu}_{nn'}$ is the inverse effective mass tensor of the
reference crystal [see Eq.\ (\ref{eq:D_total})], and
$C^{\lambda\mu\kappa}_{nn'}$ and $Q^{\lambda\mu\kappa\nu}_{nn'}$ are
the coefficients of the cubic and quartic dispersion terms in the
reference crystal [see Eqs.\ (\ref{eq:C_total}) and
(\ref{eq:Xi_total})].  These are just renormalized versions of the
quantities $\tilde{D}^{\lambda\mu}_{nn'}$,
$\tilde{C}^{\lambda\mu\kappa}_{nn'}$, and
$\tilde{Q}^{\lambda\mu\kappa\nu}_{nn'}$ defined previously in Eq.\
(\ref{eq:pi_D_C_Q}).

Likewise, the functions $W^{(2)}_{nn'} (\vect{q})$, $J^{\lambda}_{nn'}
(\vect{q})$, $M^{\lambda\mu}_{nn'} (\vect{q})$, and
$R^{\lambda\mu}_{nn'} (\vect{q})$ are all renormalized versions of
quantities defined previously.  The renormalized functions are given
explicitly in Appendix \ref{app:perturbation}.

The terms multiplying the screened nonanalytic potential $\varphi
(\vect{q})$ are derived from the Taylor series expansion and $\vect{k}
\cdot \bm{\pi}$ renormalization of the vertex function $\Lambda_{nn'}
(\vect{k}, \vect{k}')$.  The various constant coefficients (e.g.,
$\hat{R}^{\lambda\mu}_{nn'}$) are defined in the same way as the
analogous $\vect{q}$-dependent functions [e.g., $R^{\lambda\mu}_{nn'}
(\vect{q})$] given in Appendix \ref{app:perturbation}, but with
$W^{(1)}_{nn'} (\vect{q})$ replaced by the constant $\Lambda_{nn'} =
\Lambda_{nn'} (\vect{0}, \vect{0})$.

Equation (\ref{eq:em1}) can now be rearranged \cite{Leib75,Leib77} and
Fourier transformed to obtain the effective-mass Hamiltonian
\begin{widetext}
\begin{multline}
   \bar{H}_{nn'} (\vect{x}, \vect{p}) = E_n \delta_{nn'} + \{
   p_{\lambda} , \pi^{\lambda}_{nn'} (\vect{x}) \} + \{ p_{\lambda}
   p_{\mu}, D^{ \{ \lambda\mu \} }_{nn'} (\vect{x}) \} + p_{\lambda}
   p_{\mu} p_{\kappa} C^{\lambda\mu\kappa}_{nn'} + p_{\lambda} p_{\mu}
   p_{\kappa} p_{\nu} Q^{\lambda\mu\kappa\nu}_{nn'} + W^{(1)}_{nn'}
   (\vect{x}) + W^{(2)}_{nn'} (\vect{x}) \\ + \partial_{\lambda}
   Z^{\lambda}_{nn'} (\vect{x}) + \partial_{\lambda} \partial_{\mu}
   Y^{\lambda\mu}_{nn'} (\vect{x}) + \{ p_{\lambda} , \partial_{\mu}
   \Gamma^{\lambda\mu}_{nn'} (\vect{x}) \} + [ \partial_{\mu}
   \Phi^{\lambda\mu}_{nn'} (\vect{x}) ] p_{\lambda} \\ + \Lambda_{nn'}
   \varphi (\vect{x}) + \hat{\pi}^{\lambda}_{nn'} \{ p_{\lambda} ,
   \varphi (\vect{x}) \} + \hat{D}^{ \{ \lambda\mu \} }_{nn'} \{
   p_{\lambda} p_{\mu}, \varphi (\vect{x}) \} \\ +
   \hat{Z}^{\lambda}_{nn'} \partial_{\lambda} \varphi (\vect{x}) +
   \hat{Y}^{\lambda\mu}_{nn'} \partial_{\lambda} \partial_{\mu}
   \varphi (\vect{x}) + \hat{\Gamma}^{\lambda\mu}_{nn'} \{ p_{\lambda}
   , \partial_{\mu} \varphi (\vect{x}) \} +
   \hat{\Phi}^{\lambda\mu}_{nn'} [ \partial_{\mu} \varphi (\vect{x}) ]
   p_{\lambda} , \label{eq:emx}
\end{multline}
\end{widetext}
in which $\{ A, B \} = \{ A B \} = \frac12 (AB + BA)$ is the
symmetrized product, $\vect{p} = -i \nabla$ is the momentum operator,
and $\partial_{\lambda} = \partial / \partial x_{\lambda}$ acts only
on the function immediately to its right.


All of the $\vect{x}$-dependent functions are defined in terms of the
$\vect{k}$-space cutoff (\ref{eq:theta_x}).  The quantity
\begin{equation}
   \pi^{\lambda}_{nn'} (\vect{x}) = \pi^{\lambda}_{nn'} +
   J^{\lambda}_{nn'} (\vect{x}) + [J^{\lambda}_{n'n} (\vect{x})]^*
   \label{eq:pi}
\end{equation}
is the material-dependent kinetic momentum matrix for the
heterostructure, while
\begin{equation}
   D^{\lambda\mu}_{nn'} (\vect{x}) = D^{\lambda\mu}_{nn'} +
   M^{\lambda\mu}_{nn'} (\vect{x}) + [M^{\mu\lambda}_{n'n}
   (\vect{x})]^* + R^{\lambda\mu}_{nn'} (\vect{x}) \label{eq:DnnK}
\end{equation}
is half the inverse effective mass tensor for the heterostructure.
This has symmetric and antisymmetric parts:
\begin{equation}
\begin{split}
   D^{ \{ \lambda\mu \} }_{nn'} (\vect{x}) & = \tfrac12 [
   D^{\lambda\mu}_{nn'} (\vect{x}) + D^{\mu\lambda}_{nn'} (\vect{x}) ]
   , \\ 
   D^{ [ \lambda\mu ] }_{nn'} (\vect{x}) & = \tfrac12 [
   D^{\lambda\mu}_{nn'} (\vect{x}) - D^{\mu\lambda}_{nn'} (\vect{x}) ]
   ,
\end{split}
\end{equation}
although the antisymmetric part has a nonvanishing contribution only
in the presence of a magnetic field.

The functions $Z$, $Y$, $\Gamma$, and $\Phi$ in Eq.\ (\ref{eq:emx})
are all interface terms.  The first two terms
\begin{align}
   Z^{\lambda}_{nn'} (\vect{x}) & = -i \tfrac12 \big(
   J^{\lambda}_{nn'} (\vect{x}) - [J^{\lambda}_{n'n} (\vect{x})]^*
   \big) , \label{eq:Z} \\ 
   Y^{\lambda\mu}_{nn'} (\vect{x}) & = \tfrac12
   R^{\{\lambda\mu\}}_{nn'} (\vect{x}) , \label{eq:Y}
\end{align}
are renormalized versions of the $\delta$ and $\delta'$ mixing
potentials considered previously in Eqs.\ (\ref{eq:W1q}) and
(\ref{eq:W1x}).  However, the other two terms
\begin{align}
   \Gamma^{\lambda\mu}_{nn'} (\vect{x}) & = -i \big(
   M^{\{\lambda\mu\}}_{nn'} (\vect{x}) - [ M^{\{\lambda\mu\}}_{n'n}
   (\vect{x}) ]^* \big) , \label{eq:Gamma} \\
   \Phi^{\lambda\mu}_{nn'} (\vect{x}) & = i R^{[\lambda\mu]}_{nn'}
   (\vect{x}) , \label{eq:Phi}
\end{align}
were not present in Eqs.\ (\ref{eq:W1q}) and (\ref{eq:W1x}).  The term
$\Phi^{\lambda\mu}_{nn'} (\vect{x})$, which is antisymmetric in
$\lambda$ and $\mu$, is just a generalized Rashba coefficient
\cite{Rashba60,Vasko79,Wink00,Wink02,Wink03} for multi-band
Hamiltonians.  However, the term $\Gamma^{\lambda\mu}_{nn'}
(\vect{x})$, which is symmetric in $\lambda$ and $\mu$,
\cite{note:symmetric} has received little attention in the literature.
Its physical interpretation will be discussed below in Sec.\
\ref{sec:symmetry}.

Since the functions $Z$, $Y$, $\Gamma$, and $\Phi$ behave to lowest
order as step functions at an abrupt junction, the Hamiltonian
(\ref{eq:emx}) shows explicitly that these functions produce interface
terms proportional to $\delta$ or $\delta'$.

The remaining terms ($\hat{\pi}$, $\hat{Z}$, etc.)\ in Eq.\
(\ref{eq:emx}) that appear in front of $\varphi (\vect{x})$ are
defined by the obvious generalizations of Eqs.\
(\ref{eq:pi})--(\ref{eq:Phi}).  The interpretation of these terms
parallels that of the terms already discussed, except that the
contributions from $\varphi$ are not as well localized at the
interface.

Here it is worth noting that in the $GW$ ap\-prox\-i\-ma\-tion,
\cite{Hedin65,HedLund69} one has $\Lambda_{nn'} = \delta_{nn'}$ and
$\hat{\pi}^{\lambda}_{nn'} = \hat{D}^{\lambda\mu}_{nn'} = 0$, so the
potential $\varphi (\vect{x})$ does not contribute to the momentum
matrix or the effective-mass tensor.  In LDA, these simplifications
are also valid, and one has in addition $\hat{Z}^{\lambda}_{nn'} = 0$,
since the exchange-correlation potential is short-ranged.

For isovalent zinc-blende systems, the leading term in $\varphi^{(1)}$
is an octopole potential, so the contributions from the second-rank
tensors $\hat{D}$, $\hat{Y}$, $\hat{\Gamma}$, and $\hat{\Phi}$ are
negligible.  These terms are non-negligible only for the linear
interface dipole term in a heterovalent zinc-blende system (or for a
slowly varying external potential, which is not considered explicitly
here).  Likewise, $\varphi^{(2)}$ is negligible in all terms except
$\Lambda_{nn'}$.

Equation (\ref{eq:emx}) is the main result obtained in this paper.
The qualitative form of this Hamiltonian is very similar to the
Leibler Hamiltonian \cite{Leib75,Leib77} for slowly graded
heterostructures, as amended by Takhtamirov and
Volkov. \cite{TakVol97a,TakVol97b,TakVol99b} The differences are
primarily due to the use of atomic pseudopotentials (rather than a
model based on periodic bulk potentials), the inclusion of long-range
Coulomb potentials, and the use of linear response theory to simplify
the interface Hamiltonian, as discussed in Sec.\ \ref{sec:discussion}.

The explicit form of the various material parameters in Eq.\
(\ref{eq:emx}) for semiconductors with the zinc-blende structure is
given below in Sec.\ \ref{sec:symmetry}.  First, however, the
possibility of representing the material parameters as piecewise
constant is considered.

\section{Simplified mathematical representation of heterostructure 
         material properties}

\label{sec:stepfunction}

This section discusses several ways in which the mathematical
description of material properties can be simplified.  The first is to
label the materials in terms of bulk compounds (e.g., GaAs) rather
than atoms; the second is to approximate the properties of an abrupt
junction using piecewise constant material parameters with $\delta$
functions and their derivatives at interfaces.  For simplicity, only
two-dimensional isovalent systems are considered here (see Sec.\
\ref{sec:2d}).

\subsection{Transformation from atomic to bulk-crystal description}

All of the linear-response terms can be transformed immediately to a
bulk-crystal representation similar to that described above in Eq.\
(\ref{eq:Vmodel}):
\begin{subequations}
\begin{align}
   W^{(1)}_{nn'} (\vect{x}) & = \sideset{}{'} \sum_{\alpha}
   \theta^{\alpha} (\vect{x}) W^{\alpha}_{nn'} \label{eq:DV1at} \\ & =
   \sideset{}{'} \sum_{l} \theta^{l} (\vect{x}) W^{l}_{nn'} ,
   \label{eq:DV1l}
\end{align}
in which $l$ labels the different bulk materials, and [cf.\ Eq.\
(\ref{eq:theta_x_constraint})]
\begin{equation}
   \sum_{l} \theta^{l} (\vect{x}) = 0 .
\end{equation}
\end{subequations}
For example, in a GaAs/AlAs heterostructure, the linear potentials for
the two bulk media are defined by
\begin{equation}
   \begin{split}
   W^{\text{GaAs}}_{nn'} & = W^{\text{Ga}}_{nn'} 
   + W^{\text{As}}_{nn'} , \\
   W^{\text{AlAs}}_{nn'} & = W^{\text{Al}}_{nn'} 
   + W^{\text{As}}_{nn'} .
   \end{split}
\end{equation}
If the reference crystal is chosen to be GaAs, then
$W^{\text{Ga}}_{nn'} = W^{\text{As}}_{nn'} = 0$.  [This choice is made
for clarity of exposition; the final results given in Eqs.\
(\ref{eq:theta_AlAs_GaAs}), (\ref{eq:theta_InAs_GaSb}),
(\ref{eq:W_GaAs_AlAs}), and (\ref{eq:nca}) do not depend on the choice
of reference crystal.] The transformation from (\ref{eq:DV1at}) to
(\ref{eq:DV1l}) is then simply
\begin{equation}
   W^{(1)}_{nn'} (\vect{x}) = \theta^{\text{Al}} (\vect{x})
   W^{\text{Al}}_{nn'} = \theta^{\text{Al}} (\vect{x})
   W^{\text{AlAs}}_{nn'} .  \label{eq:DV_AlAs_GaAs}
\end{equation}
Hence
\begin{equation}
   \theta^{\text{AlAs}} = \theta^{\text{Al}} , \qquad
   \theta^{\text{GaAs}} = \theta^{\text{Ga}}
   . \label{eq:theta_AlAs_GaAs}
\end{equation}

For a no-common-atom system such as InAs/GaSb, four different bulk
potentials can be defined:
\begin{equation}
   \begin{split}
   W^{\text{InAs}}_{nn'} & = W^{\text{In}}_{nn'} +
   W^{\text{As}}_{nn'} , \\
   W^{\text{InSb}}_{nn'} & = W^{\text{In}}_{nn'} +
   W^{\text{Sb}}_{nn'} , \\
   W^{\text{GaAs}}_{nn'} & = W^{\text{Ga}}_{nn'} +
   W^{\text{As}}_{nn'} , \\
   W^{\text{GaSb}}_{nn'} & = W^{\text{Ga}}_{nn'} +
   W^{\text{Sb}}_{nn'} .
   \end{split}
\end{equation}
If the reference crystal is chosen to be InAs, then
$W^{\text{In}}_{nn'} = W^{\text{As}}_{nn'} = 0$.  In the atomic
description, the linear response potential is
\begin{equation}
   W^{(1)}_{nn'} (\vect{x}) =
     \theta^{\text{Ga}} (\vect{x}) W^{\text{Ga}}_{nn'} 
   + \theta^{\text{Sb}} (\vect{x}) W^{\text{Sb}}_{nn'} ,
\end{equation}
which can be rewritten (using $\theta^{\text{In}} + \theta^{\text{Ga}}
= 0$ and $\theta^{\text{As}} + \theta^{\text{Sb}} = 0$) in several
different ways, two of which are
\begin{subequations}
\begin{equation}
   W^{(1)}_{nn'} = \theta^{\text{Ga}} W^{\text{GaSb}}_{nn'} +
   (\theta^{\text{In}} + \theta^{\text{Sb}}) W^{\text{InSb}}_{nn'}
   \label{eq:DV_InSb}
\end{equation}
and
\begin{equation}
   W^{(1)}_{nn'} = \theta^{\text{Sb}} W^{\text{GaSb}}_{nn'} +
   (\theta^{\text{Ga}} + \theta^{\text{As}}) W^{\text{GaAs}}_{nn'}
   . \label{eq:DV_GaAs}
\end{equation}
\end{subequations}
Equation (\ref{eq:DV_InSb}) is useful for describing a (001)
heterojunction with an InSb-like interface:
\begin{subequations}
\begin{equation}
   \cdots \text{--As--In--As--In--Sb--Ga--Sb--Ga--} \cdots ,
   \label{eq:InSb_int}
\end{equation}
while Eq.\ (\ref{eq:DV_GaAs}) is useful for describing a
heterojunction with a GaAs-like interface:
\begin{equation}
   \cdots \text{--In--As--In--As--Ga--Sb--Ga--Sb--} \cdots .
   \label{eq:GaAs_int}
\end{equation}
\end{subequations}
In the first case one can identify the bulk functions
\begin{subequations} \label{eq:theta_InAs_GaSb}
\begin{align}
   \theta^{\text{GaSb}} & = \theta^{\text{Ga}} , &
   \theta^{\text{InAs}} & = \theta^{\text{As}} , \nonumber \\
   \theta^{\text{InSb}} & = \theta^{\text{In}} + 
      \theta^{\text{Sb}} , &
   \theta^{\text{GaAs}} & = 0 ,
\end{align}
while in the second case
\begin{align}
   \theta^{\text{GaSb}} & = \theta^{\text{Sb}} , &
   \theta^{\text{InAs}} & = \theta^{\text{In}} , \nonumber \\
   \theta^{\text{GaAs}} & = \theta^{\text{Ga}} + 
      \theta^{\text{As}} , &
   \theta^{\text{InSb}} & = 0 .
\end{align}
\end{subequations}
This type of transformation can be used for any term in the linear
response.  For the quadratic response, such a description is not
appropriate, but all contributions (including the interface dipole
potential) can be approximated as abrupt step functions, as discussed
in Secs.\ \ref{sec:quad_an} and \ref{sec:2d}.

\subsection{Piecewise constant material parameters}

\label{sec:piecewise}

The next step in simplifying the description of the material
properties is to approximate an ideal heterostructure as piecewise
constant.  As a specific example, the case of a (001) heterojunction
between semiconductors with the zinc-blende structure is considered
here.

For this case, a convenient slab-adapted \cite{Heine63,Klein81} unit
cell is defined by the basis vectors
\begin{equation}
   \vect{a}_1 = \frac{a}{2} (1, -1, 0), \quad \vect{a}_2 = \frac{a}{2}
   (1, 1, 0), \quad \vect{a}_3 = \frac{a}{2} (1, 0, 1) .
\end{equation}
Periodic boundary conditions are applied over the crystal volume
$\Omega = \vect{L}_1 \cdot (\vect{L}_2 \times \vect{L}_3)$, where
$\vect{L}_i = N_i \vect{a}_i$ and $N_i$ is an integer (thus $\Omega =
N \Omega_0$, where $N = N_1 N_2 N_3$).

The bulk properties of the heterojunction are to be represented in
terms of the periodic step function
\begin{equation}
   \Theta (\vect{x}) = \Theta(z) = \left\{
   \begin{array}{ll}
      1 , & 0 < z < \frac12 L_z \\
      0 , & -\frac12 L_z < z < 0
   \end{array}
   \right.
\end{equation}
in which $L_z = \frac12 N_3 a$ is the period in the $z$ direction, and
$\Theta (\vect{x}) = \Theta(\vect{x} + \vect{L}_i)$.  The interface
properties are to be represented by the derivatives
$\Theta_{z}(\vect{x}) = \partial \Theta / \partial z$ and
$\Theta_{zz}(\vect{x}) = \partial^2 \Theta / \partial z^2$, which are
periodic arrays of $\delta$ and $\delta'$ functions.

The first example of an actual (001) heterojunction to be considered
is a common-atom GaAs/AlAs junction.  The coordinate origin is chosen
to be an interface As atom, with unit-cell basis vectors $\bm{\tau}_a
= \vect{0}$ and $\bm{\tau}_c = \frac14 a (1,1,1)$ for anions and
cations, respectively.  The discrete function
$\theta^{\alpha}_{\vect{R}}$ for $\alpha = \text{Al}$ is therefore
\begin{equation}
   \theta^{\text{Al}}_{\vect{R}} = 
   \left\{
   \begin{array}{ll} 
      1 , & 0 \le R_3 \le \frac12 N_3 - 1 \\
      0 , & -\frac12 N_3 \le R_3 \le -1
   \end{array}
   \right.
\end{equation}
where the dimensionless coordinate $R_3$ is an integer (see Sec.\
\ref{sec:2d}).  From the Fourier transform of this function, one finds
that for small $k$,
\begin{equation}
   \theta^{\text{Al}} (\vect{k}) \simeq \Theta (\vect{k}) -
   \frac{a^2}{96} \Theta_{zz} (\vect{k}) .
\end{equation}
Upon inserting this result into Eq.\ (\ref{eq:DV_AlAs_GaAs}), one
finds that for $|z| < \frac12 L_z$, the linear band offset at a
GaAs/AlAs heterojunction can be represented as
\begin{multline}
   W^{(1)}_{nn'} (\vect{x}) = (W^{\text{AlAs}}_{nn'} -
   W^{\text{GaAs}}_{nn'}) [\Theta (z) - x] \\ - \frac{a^2}{96}
   (W^{\text{AlAs}}_{nn'} - W^{\text{GaAs}}_{nn'}) \delta' (z) ,
   \label{eq:W_GaAs_AlAs}
\end{multline}
which has been written in a general form suitable for a virtual
reference crystal $\text{(GaAs)}_{1-x}\text{(AlAs)}_x$.  Thus, if the
band offset is treated as piecewise constant, there is an additional
correction proportional to $\delta'(z)$ that has the effect of
renormalizing the matrix $Y$ in the Hamiltonian (\ref{eq:emx}).  This
merely reflects the fact that $\theta^{\text{Al}}(z)$ is a smooth
(macroscopically averaged) step function, and the difference between a
smooth step function and an abrupt step function is, to leading order,
proportional to the macroscopic average of $\delta' (z)$.

The remaining material-dependent parameters in Eq.\ (\ref{eq:emx})
can be treated using the same approach, but for these parameters the
term proportional to $\Theta_{zz}$ yields a correction of order
$(\bar{k} a)^3 \Delta \bar{V}$ or higher, which can be neglected.
Thus for an ideal GaAs/AlAs junction, all of the material parameters
in (\ref{eq:emx}) except $W^{(1)}_{nn'} (\vect{x})$ can be
replaced by abrupt step functions.

For the case of a no-common-atom InAs/GaSb junction, the situation is
more complicated.  The example considered here is the GaAs-like
junction in Eq.\ (\ref{eq:GaAs_int}), where the origin is the midpoint
of an As--Ga bond and $\bm{\tau}_c = -\bm{\tau}_a = \frac18 a
(1,1,1)$.  In this case, the small-$k$ behavior of the atomic
distribution functions is
\begin{equation}
   \begin{split}
   \theta^{\text{Ga}} (\vect{k}) & \simeq \Theta (\vect{k}) +
   \frac{a}{8} \Theta_{z} (\vect{k}) - \frac{a^2}{384} \Theta_{zz}
   (\vect{k}) , \\
   \theta^{\text{Sb}} (\vect{k}) & \simeq \Theta (\vect{k}) -
   \frac{a}{8} \Theta_{z} (\vect{k}) - \frac{a^2}{384} \Theta_{zz}
   (\vect{k}) .
   \end{split}
\end{equation}
Upon inserting these results into Eq.\ (\ref{eq:DV_GaAs}), one finds that
for $|z| < \frac12 L_z$, the linear band offset may be approximated by
\begin{subequations} \label{eq:nca}
\begin{multline}
   W^{(1)}_{nn'} (\vect{x}) = (W^{\text{GaSb}}_{nn'} -
   W^{\text{InAs}}_{nn'}) [\Theta (z) - x] \\ + \frac{a}{8} (2
   W^{\text{GaAs}}_{nn'} - W^{\text{InAs}}_{nn'} -
   W^{\text{GaSb}}_{nn'}) \delta (z) \\ - \frac{a^2}{384}
   (W^{\text{GaSb}}_{nn'} - W^{\text{InAs}}_{nn'}) \delta' (z) ,
   \label{eq:W_InAs_GaSb}
\end{multline}
which is written in a form suitable for the virtual reference crystal
$\text{(InAs)}_{1-x}\text{(GaSb)}_x$.  Note that this has the form
\begin{multline} \label{eq:nca_b}
   W^{(1)}_{nn'} (\vect{x}) = (W^{(+)}_{nn'} - W^{(-)}_{nn'}) [\Theta
   (z) - x] \\ + \frac{a}{8} (2 W^{\text{(i)}}_{nn'} - W^{(+)}_{nn'} -
   W^{(-)}_{nn'}) \delta (z) \\ - \frac{a^2}{384} (W^{(+)}_{nn'} -
   W^{(-)}_{nn'}) \delta' (z) ,
\end{multline}
\end{subequations}
in which $(+)$ and $(-)$ label the bulk materials to the right and
left of the junction, (i) labels the ``interface'' material, and the
reference crystal is $(+)_{1-x}(-)_{x}$.  This result is valid for any
no-common-atom (001) junction, regardless of the convention chosen for
the $\bm{\tau}$ vectors.  Thus for a no-common-atom junction,
$W^{(1)}_{nn'} (\vect{x})$ contains terms that renormalize both the
$Z$ and $Y$ matrices.

The $\delta(z)$ term in (\ref{eq:nca}) has a very simple physical
interpretation in which half a monolayer ($a/4$) at the interface is
occupied by GaAs instead of either InAs or GaSb.  This concept has
been used previously in the construction of envelope-function models
for electrons \cite{WSWYDO93} and phonons. \cite{Fore95b}

Note that if one replaces Sb with As in Eq.\ (\ref{eq:W_InAs_GaSb}),
the result does not reduce directly to an expression of the form
(\ref{eq:W_GaAs_AlAs}).  However, the difference is merely due to the
different choice of coordinate origin in the two cases.  To leading
order, $\Theta (z - \frac18 a) \simeq \Theta (z) - \frac18 a \delta
(z) + \frac{1}{128} a^2 \delta' (z)$, $\delta (z - \frac18 a) \simeq
\delta (z) - \frac18 a \delta' (z)$, and $\delta' (z - \frac18 a)
\simeq \delta' (z)$.  Thus, replacing Sb with As and shifting the
origin by $\frac18 a (1,1,1)$ does indeed reduce
(\ref{eq:W_InAs_GaSb}) to (\ref{eq:W_GaAs_AlAs}).

The result (\ref{eq:nca_b}) can be applied to all other
linear-response terms in the Hamiltonian (\ref{eq:emx}), but since the
$\Delta \pi (\vect{x})$ and $Z (\vect{x})$ terms are already of order
$(\bar{k} a) \Delta \bar{V}$, the $\delta'(z)$ term in
(\ref{eq:nca_b}) yields a negligible correction of order $(\bar{k}
a)^3 \Delta \bar{V}$.  Thus
\begin{multline}
   Z^{\lambda}_{nn'} (\vect{x}) = (Z^{\lambda(+)}_{nn'} -
   Z^{\lambda(-)}_{nn'}) [\Theta (z) - x] \\ + \frac{a}{8} (2
   Z^{\lambda \text{(i)}}_{nn'} - Z^{\lambda(+)}_{nn'} -
   Z^{\lambda(-)}_{nn'}) \delta (z) , \label{eq:pi_nca}
\end{multline}
and we see that $Z$ generates a correction to $Y$, while $\pi$
generates a correction to $\Gamma$.  However, the position dependence
of all remaining terms in (\ref{eq:emx}) (including the quadratic
response) can be represented as a simple step function:
\begin{equation}
   Y^{\lambda\mu}_{nn'} (\vect{x}) = (Y^{\lambda\mu(+)}_{nn'} -
   Y^{\lambda\mu(-)}_{nn'}) [\Theta (z) - x] ,
\end{equation}
since these terms are already of order $(\bar{k} a)^2 \Delta \bar{V}$.

\section{Theory of invariants}

\label{sec:symmetry}

In this section the method of invariants
\cite{Lutt56,SuzHen74,BirPik74,TrRoRa79,IvPi97} is used to construct
the explicit form of the Hamiltonian (\ref{eq:emx}) for $\Gamma$
electrons in semiconductors with the zinc-blende
structure. \cite{Parm55b,Dres55} The results are then compared with
the interface Hamiltonian given by the method of invariants for a
common-atom (001) junction, including all (linear and nonlinear)
interface terms of order $(\bar{k} a)^2 \Delta \bar{V}$ that are
permitted by symmetry.  Finally, the interface terms of order
$(\bar{k} a)\Delta \bar{V}$ are considered for the case of
$\Gamma$--$X$ coupling in GaAs/AlAs.  The results in this section make
use of the time reversal and crystal symmetry properties of the
self-energy that were derived in Ref.\ \onlinecite{Fore05a}.

\subsection{Generalized Leibler Hamiltonian}

\label{sec:Leibler}

As demonstrated above, to within terms of order $(\bar{k} a)^2 \Delta
\bar{V}$, the position dependence of the interface parameters can be
calculated entirely in terms of {\em bulk}-like matrix elements
modulated by the atomic form factors $\theta^{\alpha} (\vect{x})$.
Thus, the Hamiltonian for a zinc-blende heterostructure of arbitrary
composition can be determined by constructing invariants transforming
as $\Gamma_1$ under the symmetry operations of the $T_d$ group.  This
amounts to treating $\theta^{\alpha} (\vect{x})$ as a ``slowly
varying'' function that is an invariant of $T_d$. \cite{Leib77}

The relevant basis functions for the representations of $T_d$ are
given in Table \ref{table:Td}.
\begin{table}
\caption{\label{table:Td}Basis functions for constructing invariants
of $T_d$.  Here $\vect{P}$ is a vector operator whose components
need not commute, while $\bm{\sigma}$, $\vect{I}$, and $\vect{J}$
are angular momentum (pseudovector) operators corresponding to angular
momentum $\frac12$, 1, and $\frac32$, respectively.  Cyclic
permutations of $x$, $y$, and $z$ also yield acceptable basis
functions.}
\begin{ruledtabular}
\begin{tabular}{ll}
Rep. & Basis functions \\ \hline
$\Gamma_{1}$  &  1, $P^2$ \\
$\Gamma_{2}$  &  $J_x J_y J_z + J_z J_y J_x$ \\
$\Gamma_{12}$ &  $P_x^2 - \frac13 P^2$, $I_x^2 - \frac13 I^2$, 
$J_x^2 - \frac13 J^2$ \\
$\Gamma_{15}$ &  $P_x$, $\{ P_x P_y \}$, $\{ I_x I_y \}$,
$\{ J_x J_y \}$, $V_x \equiv \{ J_x (J_y^2 - J_z^2) \}$ \\
$\Gamma_{25}$ &  $i[P_x, P_y]$, $\sigma_x$, $I_x$, $J_x$, $J_x^3$
\end{tabular}
\end{ruledtabular}
\end{table}
The specific example to be considered here is that of the $\Gamma_8$
valence band, with $\Gamma_{15}$ and $\Gamma_6$ derivable as special
cases of the $\Gamma_8$ results.  The extension of these results to
multi-band Hamiltonians (e.g., $\Gamma_6 \oplus \Gamma_7 \oplus
\Gamma_8$) can be handled using the methods of Refs.\
\onlinecite{SuzHen74,BirPik74,TrRoRa79}, but is not considered
explicitly here.

The momentum matrix $\pi^{\lambda} (\vect{x})$ must transform as a
vector ($\Gamma_{15}$) that is odd under time reversal, which leaves
only one possibility:
\begin{equation}
   \pi^{\lambda} (\vect{x}) = V_{\lambda} \xi (\vect{x}) ,
\end{equation}
in which $V_x \equiv \{ J_x (J_y^2 - J_z^2) \}$ is a product of
angular momentum matrices $J_{\nu}$ for a particle with spin
$\frac32$, and $\xi (\vect{x})$ is some linear combination of
$\theta^{l} (\vect{x})$ functions.  The corresponding results for
$\Gamma_{15}$ and $\Gamma_6$ states are obtained by replacing $J_{\nu}
\rightarrow I_{\nu}$ (spin 1) and $J_{\nu} \rightarrow \sigma_{\nu}$
(spin $\frac12$), respectively.  This yields zero in both cases, hence
$\pi^{\lambda} (\vect{x})$ occurs only for $\Gamma_8$ and is a
relativistic effect.

For the effective-mass tensor $D^{\lambda\mu} (\vect{x})$ one obtains
the well-known result \cite{Lutt56}
\begin{multline}
   - D^{\lambda\mu} (\vect{x}) = \tfrac12 \gamma_1 (\vect{x})
   \delta_{\lambda\mu} 1 - \gamma_2 (\vect{x}) \delta_{\lambda\mu}
   (J_{\lambda}^2 - \tfrac13 J^2) \\ - \gamma_3 (\vect{x}) (1 -
   \delta_{\lambda\mu}) \{ J_{\lambda} J_{\mu} \} \\ + i
   \epsilon_{\lambda\mu\nu} [\kappa (\vect{x}) J_{\nu} +
   q(\vect{x}) J_{\nu}^3 ] , \label{eq:D_ab}
\end{multline}
in which the Luttinger parameters $\gamma_1 (\vect{x})$, $\gamma_2
(\vect{x})$, $\gamma_3 (\vect{x})$, $\kappa (\vect{x})$, and $q
(\mathbf{x})$ are all linear combinations of $\theta^{l} (\vect{x})$
functions, and $\epsilon_{\lambda\mu\nu}$ is the antisymmetric unit
tensor.  For $\Gamma_{15}$ the parameter $q$ is not independent (since
$I_{\nu}^3 = I_{\nu}$), whereas for $\Gamma_6$ only $\gamma_1$ and
$\kappa$ are independent.

The $Z$ matrix has the same symmetry as the coordinate operator (i.e.,
a vector that is even under time reversal), so it has the form
\cite{IvKaRo96}
\begin{equation}
   Z^{\lambda} (\vect{x}) = | \epsilon_{\lambda\mu\nu} | \{
   J_{\mu} J_{\nu} \} \zeta (\vect{x}) , \label{eq:Zx}
\end{equation}
where $\zeta (\vect{x})$ is a linear combination of $\theta^{l}
(\vect{x})$ functions.  This result has the same form for
$\Gamma_{15}$, but vanishes for $\Gamma_6$.  The coupling
(\ref{eq:Zx}) generates a zone-center mixing of heavy and light holes,
and was proposed independently in Refs.\ \onlinecite{AlIv92},
\onlinecite{Fore95a}, \onlinecite{Fore96}, and \onlinecite{KrVo96}.

The $Y$ interface matrix has the same symmetry as the symmetric part
of $D$, so it can be written
\begin{multline}
   Y^{\lambda\mu} (\vect{x}) = \tfrac12 \eta_1 (\vect{x})
   \delta_{\lambda\mu} 1 - \eta_2 (\vect{x}) \delta_{\lambda\mu}
   (J_{\lambda}^2 - \tfrac13 J^2) \\ - \eta_3 (\vect{x}) (1 -
   \delta_{\lambda\mu}) \{ J_{\lambda} J_{\mu} \} .
\end{multline}
This term has not been studied previously for $\Gamma_8$ states,
although the corresponding term for $\Gamma_6$ electrons (a direct
analog of the Darwin term from the Dirac equation) is well
known. \cite{Leib75,Leib77,Young89}

The $\Gamma$ matrix has the form
\begin{equation}
   \Gamma^{\lambda\mu} (\vect{x}) = | \epsilon_{\lambda\mu\nu}
   | V_{\nu} \chi (\vect{x}) , \label{eq:Gab}
\end{equation}
in which $\chi (\vect{x})$ is a linear combination of $\theta^{l}
(\vect{x})$ functions.  The closely related coupling
$\hat{\Gamma}^{\lambda\mu}$ arising from an external electric field
was proposed recently in Ref.\ \onlinecite{Gan02b} as a possible
mechanism for the linear and circular photogalvanic effects.  This
term is relativistic in origin and does not occur for $\Gamma_{15}$ or
$\Gamma_6$.

The matrix $\Phi$ has the same symmetry as the antisymmetric part of
$D$, except that it is odd rather than even under time reversal.  It
can therefore be written as
\begin{equation}
   \Phi^{\lambda\mu} (\vect{x}) = \epsilon_{\lambda\mu\nu}
   [\kappa' (\vect{x}) J_{\nu} + q' (\vect{x}) J_{\nu}^3 ]
   . \label{eq:Phi_ab}
\end{equation}
For $\Gamma_6$ electrons (where $q'$ is not independent) this is the
analog of the Rashba spin-splitting effect for low-symmetry bulk
semiconductors, \cite{Rashba60} which occurs also in heterostructures
of cubic semiconductors \cite{Leib77,Vasko79,Bast81,Bast82} due to the
reduced symmetry at a surface or interface.  The corresponding
Hamiltonian for $\Gamma_8$ states was proposed in Ref.\
\onlinecite{Fore93}, and has received renewed attention as the
valence-band Rashba coupling \cite{Wink00,Wink02,Wink03,BernZhang05a}
in recent years.

The coefficients of the various terms involving $\varphi$ in Eq.\
(\ref{eq:emx}) have the same form as those already given.  The
symmetry restrictions on the Taylor series expansions for $n^{(1)}$
and $n^{(2)}$ were given in Ref.\ \onlinecite{Fore05a}.

\subsection{GaAs/AlAs (001) heterojunction}

The significance of these results is now investigated by comparing
them with all interface terms of order $(\bar{k} a)^2 \Delta \bar{V}$
allowed by symmetry for a GaAs/AlAs (001) heterojunction, which has
the point group $C_{2v}$.  To better understand this comparison, it is
helpful to begin by studying the symmetry properties of the linear and
quadratic response in this system.

The starting point is the observation that although the point group of
the heterojunction is $C_{2v}$ (with the coordinate origin at an
interface As atom), if the reference crystal is chosen to be the
virtual crystal Al$_{0.5}$Ga$_{0.5}$As, then the perturbation due to
the ionic pseudopotentials has a higher symmetry: it transforms
according to the representation $X_3$ of the $D_{2d}$ group.  In other
words, it transforms according to the identity representation
$\Delta_1$ of the $C_{2v}$ group (none of whose operations change the
$z$ coordinate), but in addition it has odd parity with respect to
those elements of $D_{2d}$ that change the sign of $z$.  This occurs
because the atomic mole fraction $\theta^{\alpha }_{\vect{R}}$ itself
transforms as $X_3 (D_{2d})$, whereas the atomic perturbation $\Delta
v^{\alpha}_{\text{ion}} (\vect{x}, \vect{x}')$ has the site symmetry
$\Gamma_1 (T_d)$ or $X_1 (D_{2d})$.  Therefore, $\Delta V_{\text{ion}}
(\vect{x}, \vect{x}')$ transforms as $X_1 \otimes X_3 = X_3$ under the
operations of $D_{2d}$.

To determine the behavior of the screened potential $\Delta V
(\vect{x}, \vect{x}')$, note that since the vertex functions
\cite{Fore05a} $\Gamma^{(1)}$ and $\Gamma^{(2)}$ of the reference
crystal transform as $\Gamma_1 (T_d)$, the linear response $V^{(1)}
(\vect{x}, \vect{x}')$ transforms as $X_1 \otimes X_3 = X_3$, while
the quadratic response $V^{(2)} (\vect{x}, \vect{x}')$ transforms as
$X_1 \otimes X_3 \otimes X_3 = X_1$.  (In general, all odd-order terms
in the response transform as $X_3$, while all even-order terms
transform as $X_1$.)  Thus, the total response $\Delta V (\vect{x},
\vect{x}') = V^{(1)} (\vect{x}, \vect{x}') + V^{(2)} (\vect{x},
\vect{x}')$ transforms as neither $X_3$ nor $X_1$, but as $\Delta_1
(C_{2v})$.

If the quadratic response is of the same order as the linear response,
then one must include all possible invariants of $C_{2v}$ that are of
order $(\bar{k} a)^2 \Delta \bar{V}$ when constructing the interface
Hamiltonian.  Basis functions for $D_{2d}$ and $C_{2v}$ are given in
Tables \ref{table:D2d} and \ref{table:C2v}, respectively; note that
both $X_1$ and $X_3$ are compatible with $\Delta_1$.
\begin{table}
\caption{\label{table:D2d}Basis functions for $D_{2d}$.  The
components of $\bm{\sigma}$, $\vect{I}$, and $\vect{B} = i c (\vect{P}
\times \vect{P})$ transform as those of $\vect{J}$.}
\begin{ruledtabular}
\begin{tabular}{ll}
Rep. & Basis functions \\ \hline
$X_{1}$ &  1, $P_x^2 + P_y^2$, $P_z^2$, $J_x^2 + J_y^2$, $J_z^2$,
           $P_z \{ J_x J_y \}$, $P_z V_z$, \\
        &  $P_x J_x - P_y J_y$, $P_x J_x^3 - P_y J_y^3$, 
           $P_x V_x + P_y V_y$, \\
        &  $P_x \{J_y J_z\} + P_y \{J_z J_x\}$ \\
$X_{2}$ &  $P_x^2 - P_y^2$, $J_x^2 - J_y^2$, $P_z J_z$, $P_z J_z^3$,
           $P_x J_x + P_y J_y$, \\
        &  $P_x J_x^3 + P_y J_y^3$, $P_x V_x - P_y V_y$,
           $P_x \{J_y J_z\} - P_y \{J_z J_x\}$ \\
$X_{3}$ &  $P_z$, $\{ P_x P_y \}$, $\{ J_x J_y \}$, $V_z$, $P_z J_z^2$,
           $P_x J_y - P_y J_x$, \\
        &  $P_x J_y^3 - P_y J_x^3$, $P_x V_y + P_y V_x$, 
           $P_x \{J_x J_z\} + P_y \{J_y J_z\}$ \\
$X_{4}$ &  $J_z$, $J_z^3$, $P_z (J_x^2 - J_y^2)$,
           $P_x J_y + P_y J_x$, \\
        &  $P_x J_y^3 + P_y J_x^3$, $P_x V_y - P_y V_x$, 
           $P_x \{J_x J_z\} - P_y \{J_y J_z\}$ \\
$X_{5}$ &  $(P_x, P_y)$, $(\{ P_y P_z \}, \{ P_z P_x \})$, 
           $(J_x, -J_y)$, $(J_x^3, -J_y^3)$, \\
        &  $(V_x, V_y)$, $(\{J_y J_z \}, \{ J_z J_x \} )$, 
           $(P_z J_y, -P_z J_x)$, \\
        &  $(P_z J_y^3, -P_z J_x^3)$, 
           $(P_z \{J_z J_x\}, P_z \{J_z J_y\})$, $(P_z V_y, P_z V_x)$, \\
        &  $(P_y J_z, -P_x J_z)$, $(P_x J_z^2, P_y J_z^2)$, 
           $(P_y J_z^3, -P_x J_z^3)$, \\
        &  $(P_y V_z, P_x V_z)$, $(P_y \{J_x J_y\}, P_x \{J_x J_y\})$
\end{tabular}
\end{ruledtabular}
\end{table}
\begin{table}
\caption{\label{table:C2v}Basis functions for $C_{2v}$.  The
components of $\bm{\sigma}$, $\vect{I}$, and $\vect{B} = i c (\vect{P}
\times \vect{P})$ transform as those of $\vect{J}$.}
\begin{ruledtabular}
\begin{tabular}{ll}
Rep. & Basis functions \\ \hline
$\Delta_{1}$ &  1, $P_z$, $\{ P_x P_y \}$, $P_x^2 + P_y^2$, $P_z^2$,
                $\{ J_x J_y \}$, $J_x^2 + J_y^2$, $J_z^2$, 
                $J_z^3$, \\ & $V_z$ \\
$\Delta_{2}$ &  $P_x^2 - P_y^2$, $J_z$, $J_x^2 - J_y^2$ \\
$\Delta_{3}$ &  $P_x + P_y$, $\{ P_z (P_x + P_y) \}$, $J_x - J_y$,
                $\{ J_z (J_x + J_y) \}$, \\
             &  $J_x^3 - J_y^3$, $V_x + V_y$ \\
$\Delta_{4}$ &  $P_x - P_y$, $\{ P_z (P_x - P_y) \}$, $J_x + J_y$,
                $\{ J_z (J_x - J_y) \}$, \\
             &  $J_x^3 + J_y^3$, $V_x - V_y$
\end{tabular}
\end{ruledtabular}
\end{table}
All interface invariants of $C_{2v}$ that are (1)~hermitian,
(2)~time-reversal invariant, and (3)~of order $(\bar{k} a)^2 \Delta
\bar{V}$ or less are listed in Table \ref{table:intC2v}.
\begin{table}
\caption{\label{table:intC2v}Terms in the (001) GaAs/AlAs interface
Hamiltonian constructed from invariants of $C_{2v}$.}
\begin{ruledtabular}
\begin{tabular}{lccc}
Term & $D_{2d}$ & $\Gamma_1 (T_d)$ & Origin \\ \hline
$1 \delta(z)$                        & $X_1$ & no  & --- \\
$(J_z^2 - \frac13 J^2) \delta(z)$    & $X_1$ & no  & --- \\
$\{ J_x J_y \} \delta(z)$            & $X_3$ & yes & $Z^{\lambda}$ \\
$1 \delta'(z)$                       & $X_3$ & yes & $Y^{\lambda\mu}$ \\
$(J_z^2 - \frac13 J^2) \delta'(z)$   & $X_3$ & yes & $Y^{\lambda\mu}$ \\
$\{ J_x J_y \} \delta'(z)$           & $X_1$ & no  & --- \\
$V_z \{ P_z \delta(z) \}$            & $X_1$ & no  & --- \\
$(P_x V_x + P_y V_y) \delta(z)$      & $X_1$ & no  & --- \\
$(P_x V_y + P_y V_x) \delta(z)$      & $X_3$ & yes & $\Gamma^{\lambda\mu}$ \\
$(P_x J_x - P_y J_y) \delta(z)$      & $X_1$ & no  & --- \\
$(P_x J_y - P_y J_x) \delta(z)$      & $X_3$ & yes & $\Phi^{\lambda\mu}$ \\
$(P_x J_x^3 - P_y J_y^3) \delta(z)$  & $X_1$ & no  & --- \\
$(P_x J_y^3 - P_y J_x^3) \delta(z)$  & $X_3$ & yes & $\Phi^{\lambda\mu}$
\end{tabular}
\end{ruledtabular}
\end{table}

The second column in this table lists the symmetry of each term under
the operations of $D_{2d}$.  The third column indicates whether the
term can be constructed as an invariant of $T_d$, treating $\Theta(z)$
as an invariant of $T_d$.  (As discussed above, the latter approach
can be used for all linear-response interface terms.)  Note that in
the $T_d$ symmetry analysis, $\delta (z) = d\Theta / dz$ transforms as
$z$ ($\Gamma_{15}$) and $\delta'(z)$ transforms as $z^2$ ($\Gamma_1$
or $\Gamma_{12}$), whereas in the $D_{2d}$ symmetry analysis, $\delta
(z)$ transforms as $z^2$ ($X_1$) and $\delta' (z)$ transforms as $z$
($X_3$).

In agreement with the general symmetry properties of $V^{(1)}$ and
$V^{(2)}$ derived earlier, Table \ref{table:intC2v} shows that all
linear interface terms [derivable from $\Gamma_1 (T_d)$] transform as
$X_3 (D_{2d})$.  The remaining terms that transform as $X_1$ all
originate in the quadratic or higher-order response.  Thus, if one
accounts for the smallness of the quadratic response (Sec.\
\ref{sec:est_quad}), seven out of the thirteen possible interface
invariants for (001) GaAs/AlAs heterojunctions can be omitted because
they are actually of order $(\bar{k} a)^3 \Delta \bar{V}$ or higher.
This represents a major simplification over the general case in which
all response terms are of the same order.

For a no-common-atom heterojunction, the results of Sec.\
\ref{sec:Leibler} remain valid (as they depend only on the bulk
symmetry), but the linear response now has only the symmetry $\Delta_1
(C_{2v})$.  Thus the terms labeled $X_1$ in Table \ref{table:intC2v}
now arise (in general) even in linear-response theory, due to the
renormalization effects described above in Eqs.\ (\ref{eq:nca}) and
(\ref{eq:pi_nca}).  For example, the $\delta(z)$ term in Eq.\
(\ref{eq:nca}) generates the term in the first row of Table
\ref{table:intC2v}.  However, the term in the second row of Table
\ref{table:intC2v} is still zero (within linear response), because
$W^{(1)}_{nn'} (\vect{x})$ in Eq.\ (\ref{eq:nca}) couples only states
of the same symmetry, with $W^{(1)}_{nn'} (\vect{x}) = \delta_{nn'}
W^{(1)}_{nn} (\vect{x})$ and $W^{(1)}_{nn} (\vect{x}) = W^{(1)}_{n'n'}
(\vect{x})$ whenever $E_n = E_{n'}$.

\subsection{$\Gamma$--$X$ coupling}

\label{sec:GammaX}

If one extends the above analysis to the case of intervalley
$\Gamma$--$X$ coupling at a (001) GaAs/AlAs junction, it is
immediately apparent (because the $Z$ matrix has the symmetry of a
coordinate matrix) that the linear response produces $\delta(z)$
coupling between the bands
\begin{equation}
   \Gamma_1 \text{--} X_{3z}, \qquad X_{1z} \text{--} X_{3z}, \qquad
   X_{3x} \text{--} X_{3y} , \label{eq:X3}
\end{equation}
but no $\delta(z)$ coupling between the bands
\begin{equation}
   \Gamma_1 \text{--} X_{1z}, \qquad X_{1x} \text{--} X_{1y}
   . \label{eq:X1}
\end{equation}
Coupling of the latter type only occurs in the quadratic response, or
in the linear response from terms proportional to $\delta' (z)$ or $i
\{ P_{z} \delta(z) \}$.  These conclusions agree with those of Ref.\
\onlinecite{Fore98b}, which were derived from a model potential
constructed from a linear superposition of atomic-like
pseudopotentials.

Note, however, that these results hold only for an ideal
heterojunction.  If interdiffusion of Ga and Al atoms breaks the $X_3$
symmetry of the linear response, then there will be $\delta (z)$
coupling between the bands (\ref{eq:X1}) due to terms similar to those
derived in Eq.\ (\ref{eq:nca}).

\section{Discussion}

\label{sec:discussion}

Since a summary of the main results of this paper has already been
given in Sec.\ \ref{sec:summary}, this section will be limited to a
discussion of the differences between the present theory and previous
envelope-function theories that have appeared in the literature.  The
principle differences are that (1)~the present Hamiltonian is
constructed from atomic pseudopotentials rather than the periodic
potential of a bulk crystal; (2)~the present theory is
self-consistent, accounting fully for electron-electron Coulomb
interactions; and (3)~the present approach uses linear and quadratic
response theory to simplify the functional form of the heterostructure
Hamiltonian.

Starting with point (1), previous envelope-function theories
\cite{Leib75,Leib77,Burt88a,Burt92,Burt94c,Burt99,Young89,CuVH91,%
KaTi91,KaKr96a,KaKr96b,Fore93,Fore95a,Fore96,Fore98b,Fore01b,BuFo99,%
Klip98,TakVol97a,TakVol97b,TakVol99a,TakVol99b,TakVol00a,%
Cortez00a,Cortez01c,SmMa86,SmMa90,MaSm90a,Trze88b,AnWaAk89,%
CuVH92,CuVH93a,CuVH93b,AvSi93a,AvSi94,GrKaCh94} were based on model
potentials of the form (\ref{eq:Vmodel}), which is a linear
combination of periodic bulk potentials multiplied by step-like
functions $\theta^{l} (\vect{x})$.  The form of $\theta^{l}
(\vect{x})$ near an interface was either specified as part of the
model (e.g., abrupt step function) or treated as unknown---and in
principle unknowable, at least within the model potential approach.

Although the present theory led to a similar expression
(\ref{eq:DV1l}), this was a {\em derived} result based on linear
response theory, in which the functions $\theta^{l} (\vect{x})$ have a
known form determined by the distribution $\theta^{\alpha}_{\vect{R}}$
of atoms in the heterostructure.  The potential (\ref{eq:DV1l})
includes contributions from the ``interface'' materials (e.g., GaAs
and InSb in an InAs/GaSb heterostructure), which are often omitted in
the model-potential approach.  Furthermore, for the model potential,
the coefficients of the $\delta$ terms in the Hamiltonian depend on
(i)~the values of the {\em bulk} potentials {\em at} the reciprocal
lattice vectors $\vect{G}$; and (ii)~whether the functions $\theta^{l}
(\vect{x})$ are smooth (zero outside the Brillouin zone) or sharp,
with some $\delta$ terms vanishing in the smooth case.
\cite{TakVol97a,TakVol97b,TakVol99a,TakVol99b,TakVol00a} In the
present theory these coefficients are determined not by $\theta^{l}
(\vect{x})$, but instead by the properties of the linear {\em atomic}
pseudopotentials in a {\em finite neighborhood} of each $\vect{G}$.

For point (2), the present theory includes long-range multipole
Coulomb potentials (because the self-energy is nonanalytic at
$\vect{q} = \vect{G}$), which were not considered in previous studies.
These terms have no qualitative effect in two-dimensional systems
(other than to renormalize the band offsets), but they are present in
quantum wires and dots, even for isovalent heterostructures.

Finally, for point (3), if the functions $\theta^{l} (\vect{x})$ are
treated as unknown in the model potential approach, one has no way of
knowing how large the various interface terms are, and all terms
permitted by symmetry should in principle be included.
\cite{TakVol97a,TakVol97b,TakVol99a,TakVol99b,TakVol00a} However,
since the linear response is dominant (for typical heterostructures)
in a first-principles theory, one can eliminate all nonlinear
interface band-mixing terms, thereby simplifying the Hamiltonian
considerably.

Aside from these differences, the qualitative form of the Hamiltonian
derived here is very similar to those derived by Leibler
\cite{Leib75,Leib77} for slowly graded heterostructures and
Takhtamirov and Volkov
\cite{TakVol97a,TakVol97b,TakVol99a,TakVol99b,TakVol00a} for abrupt
heterostructures.  The basic structure of these Hamiltonians is
similar because the same type of perturbation theory
\cite{LuKo55,BirPik74,Wink03} was used in their derivation.

Other differences arise in theories using different approximation
techniques.  The well-known theory of Burt \cite{Burt92} is based on
the Luttinger-Kohn rep\-re\-sen\-ta\-tion \cite{LuKo55,Leib75} with an
energy-dependent approximation similar to L\"owdin perturbation theory
\cite{Lowd51} used to eliminate the coupling to remote bands.
Estimation of the error involved in this approximation shows that
Burt's general theory \cite{Burt92} includes all terms of order
$(\bar{k} a) \Delta \bar{V}$, but retains some terms of order
$(\bar{k} a)^2 \Delta \bar{V}$ while omitting others.  In particular,
his results give an essentially correct (within the model potential
approach) description of the $\delta$-function term $Z$ and the Rashba
term $\Phi$, but omit the contributions from remote-band coupling to
the terms $Y$ and $\Gamma$.

In Ref.\ \onlinecite{Fore93}, I used Burt's theory, but introduced an
additional approximation (also adopted in Refs.\ \onlinecite{Burt99}
and \onlinecite{Fore01b}) whereby the energy eigenvalue in the
denominator of the energy-dependent effective mass was replaced with
the position-dependent energy of the bulk valence band maximum.  This
has the effect of replacing the Rashba coefficient in Eq.\
(\ref{eq:Phi}) with $i D^{[\lambda\mu]}_{nn'} (\vect{x})$.  In other
words, the Rashba parameters $\kappa'$ and $q'$ of Eq.\
(\ref{eq:Phi_ab}) were replaced by the Luttinger parameters $\kappa$
and $q$ of Eq.\ (\ref{eq:D_ab}).  This simplifies applications of the
theory because $\kappa$ and $q$ are known from bulk magnetoabsorption
measurements, \cite{PiBr66,PiGr69} and $\kappa$ can also be estimated
from $\gamma_1$, $\gamma_2$, and $\gamma_3$. \cite{PiBr66,Fore93}
However, since the true Rashba parameters $\kappa'$ and $q'$ cannot be
determined from bulk measurements, this simplification will produce
values that have the correct order of magnitude but are quantitatively
incorrect.

A similar approximation was used in Refs.\ \onlinecite{KrVo96},
\onlinecite{Cortez00a}, and \onlinecite{Cortez01c} to estimate the
$\delta$-function mixing parameter for light and heavy holes [see
Table \ref{table:intC2v} and Eq.\ (\ref{eq:Zx})] from the valence-band
offset.  The original proposal was based on the $H_{BF}$ model,
\cite{KrVo96} in which the zone-center $\Gamma_{15}$ Hamiltonian is
written as a sum of operators $B = \frac12 + \{ I_x I_y \}$ and $F =
\frac12 - \{ I_x I_y \}$ multiplied by position-dependent bulk
valence-band energies $E_v$.  For a (001) GaAs/AlAs heterojunction
this yields the $\Gamma_{15}$ interface Hamiltonian
\begin{equation}
    \frac{a}{4} \{ I_x I_y \} (E_v^{\text{GaAs}} - E_v^{\text{AlAs}})
    \delta (z) , \label{eq:HBF1}
\end{equation}
whereas for an InAs/GaSb junction with a GaAs-like interface one
obtains
\begin{equation}
    \frac{a}{4} (1 - 2 \{ I_x I_y \}) [ E_v^{\text{GaAs}} - \tfrac12
    (E_v^{\text{InAs}} + E_v^{\text{GaSb}}) ] \delta (z) .
    \label{eq:HBF2}
\end{equation}
The corresponding results for $\Gamma_8$ are given by the substitution
$\{ I_x I_y \} \rightarrow \frac13 \{ J_x J_y \}$. \cite{Lutt56} Now
the diagonal term [proportional to $1 \, \delta(z)$] in Eq.\
(\ref{eq:HBF2}) is identical (within linear response) to the
$\delta(z)$ term derived above in Eq.\ (\ref{eq:nca}).  However, the
valence-band {\em mixing} term derived in Eqs.\ (\ref{eq:Z}) and
(\ref{eq:Zx}) is not related to the valence-band offset.  Therefore,
although the model of Refs.\ \onlinecite{KrVo96},
\onlinecite{Cortez00a}, and \onlinecite{Cortez01c} yields a coupling
of the correct symmetry and order of magnitude, its numerical value is
not reliable. \cite{Fore99} The same conclusion was reached in Ref.\
\onlinecite{MagZun00}.

In Ref.\ \onlinecite{Fore96}, I proposed a smooth model po\-ten\-tial
\cite{note:Vmodel} for abrupt heterostructures that yields a
valence-band mixing determined chiefly by the change in Bloch
functions at an interface.  This result is of course not valid for
general model po\-ten\-tials \cite{Fore96,Fore98b} (despite claims to
the contrary\cite{KaKr96a,KaKr96b}), and it is not supported by the
present work.  However, in Ref.~\onlinecite{TakVol99b} it was argued
that the valence-band mixing generated by Bloch-function differences
is negligible in comparison to that generated by a discontinuous model
potential.  Although the latter does not appear in the present
first-principles theory, its analog is the $Z$ term shown in Eq.\
(\ref{eq:W1x}).  The former contribution is given by the term
proportional to $W_{in'}^{(1)} (\vect{q})$ in Eqs.\ (\ref{eq:J_total})
and (\ref{eq:Z}).  Inspection of these terms shows that both are of
order $(\bar{k} a) \Delta \bar{V}$.  Therefore, neither the model of
Ref.~\onlinecite{Fore96} [which omits the interface contributions from
Eq.\ (\ref{eq:W1q})] nor the model of Refs.\
\onlinecite{TakVol97b,TakVol99a,TakVol99b} (which omits the
contribution from Bloch-function differences) gives a correct value
for the valence-band mixing coefficient.

In Refs.\ \onlinecite{Klip99,Klip00,Klip01}, \onlinecite{Klip98}, and
\onlinecite{TakVol00a} it was proposed that intervalley
$\Gamma_1$--$X_{1z}$ and $X_{1x}$--$X_{1y}$ coupling should be
proportional to $\delta(z)$ and of the same order of magnitude as
$\Gamma_1$--$X_{3z}$, $X_{1z}$--$X_{3z}$, and $X_{3x}$--$X_{3y}$
coupling.  However, as shown above in Sec. \ref{sec:GammaX}, the
former types of coupling should generally be substantially weaker than
the latter for ideal interfaces.  The same conclusion was reached in
Ref.\ \onlinecite{WaZu97}, where it was shown that despite the very
small magnitude of the {\em direct} $\Gamma_1$--$X_{1z}$ coupling,
there is a net {\em effective} $\Gamma_1$--$X_{1z}$ coupling (as
revealed by the anticrossing of superlattice subbands in Fig.\ 1 of
Ref.\ \onlinecite{WaZu97}) generated by $\Gamma_1$--$X_{3z}$,
$X_{1z}$--$X_{3z}$, and $\vect{k} \cdot \vect{p}$ couplings that is
only a factor of two or three smaller than the net
$\Gamma_1$--$X_{3z}$ coupling.

However, in Refs.\ \onlinecite{Klip99,Klip00,Klip01} it was found that
$X_x$--$X_y$ mixing experiments could not adequately be explained
without including $\delta$-function $X_{1x}$--$X_{1y}$ coupling of
comparable magnitude to $X_{3x}$--$X_{3y}$ coupling.  Nevertheless,
since Refs.\ \onlinecite{Klip99,Klip00,Klip01} did not test the effect
of $\delta'(z)$ $X_{1x}$--$X_{1y}$ coupling, it is not yet clear that
their experiments require such a large $\delta(z)$ coupling.  If
$\delta' (z)$ or $i \{ P_{z} \delta(z) \}$ coupling is unable to
explain their results, then their data may indicate the presence of
nonideal interfaces, as discussed in Sec.\ \ref{sec:GammaX}.  Another
possible explanation would be spin-orbit coupling, which generates a
direct $\delta(z)$ coupling between $X_{6x}$ and $X_{6y}$ even for
ideal interfaces.

Two important physical effects not considered in this paper are the
contributions from alloying (beyond the virtual crystal approximation)
and from strain (in the bulk or at an interface).  The former effect
is in principle already encompassed by the present formalism (for
lattice-matched systems in which linear response is dominant),
although for practical applications one would need to perform a
detailed analysis of the significance of different intervalley mixing
effects. \cite{WaZu97} The contribution from strain would require a
nontrivial extension of this theory, but since an {\em ab initio}
linear-response approach to strain has already been successfully
applied to In$_{0.53}$Ga$_{0.47}$As/InP, \cite{BaReBaPe89,PBBR90}
Si/Ge, \cite{ColResBar91,PerBar94} and InAs/GaSb \cite{MonPer96}
within LDA, the general perturbation scheme used in this paper should
work for lattice-mismatched systems also.

\begin{acknowledgments}
I am grateful to Eduard Takhtamirov and Roland Winkler for valuable
comments on an earlier version of the manuscript.  This work was
supported by Hong Kong RGC Grant HKUST6139/00P and UGC Grant
HIA03/04.SC02.
\end{acknowledgments}

\appendix

\section{Imaginary part of the self-energy}

\label{app:imag_sigma}

Luttinger \cite{Lutt61} has shown for homogeneous systems that
$\Sigma^{\text{(i)}}$ vanishes at the Fermi surface to all orders in
perturbation theory.  The generalization of his result to the case of
inhomogeneous systems is $\Sigma^{\text{(i)}}(\omega \pm i 0^{+}) =
\mp J(\omega)$, where $\omega$ is real and
\cite{NegOrl98_p254,Farid99}
\begin{equation}
   J(\omega) = \left\{
   \begin{array}{ll}
      C_{+} (\omega - \mu_{+})^2 , & \omega \ge \mu_{+} \\
      0, & \mu_{-} \le \omega \le \mu_{+} \\
      C_{-} (\omega - \mu_{-})^2 , & \omega \le \mu_{-}
   \end{array}
   \right.
\end{equation}
for $\omega$ near $\mu_{\pm} \equiv \mu \pm \frac12 E_{\text{g}}$.
Here $\mu$ is the chemical potential (in the limit of zero
temperature), $E_{\text{g}}$ is the energy gap, and $C_{\pm} \ge 0$ is
independent of $\omega$.  The magnitude of $C_{\pm}$ may be estimated
from the calculations of Quinn and Ferrell \cite{QuiFer58} for a
degenerate homogeneous electron gas.  Their results may be expressed
as $\langle C_{\pm} \rangle = 1 / E_{\text{i}}$, where
\cite{QuiFer58,Call91_p820}
\begin{equation}
   E_{\text{i}} = \frac{256}{\sqrt{3} \pi^2}
   \frac{\epsilon_{\text{F}}^2}{\omega_{\text{p}}} ,
\end{equation}
in which $\epsilon_{\text{F}}$ is the Fermi energy and
$\omega_{\text{p}}$ is the plasma frequency.  If one treats
semiconductors with the diamond or zinc-blende structure as a
homogeneous gas with a density of eight valence electrons per
primitive unit cell, one obtains values of $E_{\text{i}} \simeq 126$
eV for GaAs and $E_{\text{i}} \simeq 140$ eV for Si.  This is very
large in comparison to the energy gap and suggests that
$\Sigma^{\text{(i)}}$ will have a negligible influence on the band
structure in typical semiconductor heterostructures (although it has
an important qualitative effect in producing a finite quasiparticle
lifetime).

Since a real semiconductor is not a degenerate homogeneous electron
gas, one may question whether this estimate is reliable.  However, the
calculations of Fleszar and Hanke \cite{FleHan97} for Si yield
$E_{\text{i}} \agt 100$ eV (estimated from Fig.\ 3 of Ref.\
\onlinecite{FleHan97}), with some asymmetry between electrons and
holes.  Thus it is reasonable in heterostructures to assume that
$E_{\text{i}} \gg \bar{E}_{\text{g}}$.  Since $(\omega - \mu_{\pm})^2$
is of order $(\Delta \bar{V})^2$, this means that
$\Sigma^{\text{(i)}}$ is negligible under the present perturbation
scheme.  Thus it will, for the most part, not be considered explicitly
in the envelope-function equations derived here.  However, the leading
contribution from $\Sigma^{\text{(i)}}$ is noted in Eq.\
(\ref{eq:sig_nn}) so that it may be included if desired.

\section{Cutoff function}

\label{app:cutoff}

This appendix considers possible definitions of the Brillouin zone
cutoff function $B (\vect{k})$ introduced in Eq.\ (\ref{eq:theta_x}).
The most obvious choice [from Eq.~(\ref{eq:four_theta_a})] would be
\begin{equation}
   B(\vect{k}) = \left\{
   \begin{array}{ll}
      1 , & \vect{k} \in \Omega_0^* \\
      0 , & \vect{k} \notin \Omega_0^*
   \end{array}
   \right.  \label{eq:S0}
\end{equation}
which was used previously in Refs.\ \onlinecite{Burt88a} and
\onlinecite{Burt92}.  However, the sharp cutoff at the boundary of
$\Omega_0^*$ has the undesirable effect of producing Gibbs
oscillations in $\vect{x}$ space.  Other possibilities that eliminate
this problem are $B(\vect{k}) = [1 + \beta (k a)^2] \exp [-\beta (k
a)^2]$ and $B(\vect{k}) = \exp [-\beta (k a)^4]$, in which $\beta$ is
some number of order 1.  These choices of $B(\vect{k})$ are smooth,
spherically symmetric, and introduce a negligible error of order
$(\bar{k} a)^4 \Delta \bar{V}$.

For some applications one need not introduce any cutoff function at
all, as an alternative power-series approximation for $\theta^{\alpha}
(\vect{k})$ is often more convenient.  This approximation is discussed
in Sec.\ \ref{sec:piecewise}.

\begin{widetext}

\section{Renormalized effective-mass parameters}

\label{app:perturbation}

This appendix presents some details of the perturbation theory used in
Sec.\ \ref{sec:perturbation}.  For states $m, m' \in A$, the effective
Hamiltonian $\bar{H}$ is given by \cite{BirPik74,Wink03}
\begin{multline}
   \bar{H}_{mm'} = H_{mm'} + \frac12 \sum_{i}^{B} H_{mi}' H_{im'}'
   \biggl( \frac{1}{E_{mi}} + \frac{1}{E_{m'i}} \biggr) + \frac12
   \sum_{i}^{B} \sum_{j}^{B} H_{mi}' H_{ij}' H_{jm'}' \biggl(
   \frac{1}{E_{mi} E_{mj}} + \frac{1}{E_{m'i} E_{m'j}} \biggr) \\ -
   \frac12 \sum_{i}^{B} \sum_{m''}^{A} \biggl( \frac{H_{mm''}'
   H_{m''i}' H_{im'}' }{E_{mi} E_{m''i}} + \frac{H_{mi}' H_{im''}'
   H_{m''m'}'}{E_{m'i} E_{m''i}} \biggr) + \order \bigl( (H')^4 \bigr)
   , \label{eq:bp138}
\end{multline}
in which $E_{mi} = E_m - E_i$.  This expression can be used to derive
most of the terms in the effective-mass Hamiltonian (\ref{eq:em1}).
The coefficients of the quadratic, cubic, and quartic dispersion terms
of the reference crystal are given by
\begin{equation}
   D^{\lambda\mu}_{nn'} = \tilde{D}^{\lambda\mu}_{nn'} + \frac12
   \sum_{i}^{B} \pi^{\lambda}_{ni} \pi^{\mu}_{in'} \biggl(
   \frac{1}{E_{ni}} + \frac{1}{E_{n'i}} \biggr) , \label{eq:D_total}
\end{equation}
\begin{multline}
   C^{\lambda\mu\kappa}_{nn'} = \tilde{C}^{\lambda\mu\kappa}_{nn'} +
   \frac12 \sum_{i}^{B} (\pi^{\lambda}_{ni}
   \tilde{D}^{\mu\kappa}_{in'} + \tilde{D}^{\lambda\mu}_{ni}
   \pi^{\kappa}_{in'}) \biggl( \frac{1}{E_{ni}} + \frac{1}{E_{n'i}}
   \biggr) + \frac12 \sum_{i}^{B} \sum_{j}^{B} \pi^{\lambda}_{ni}
   \pi^{\mu}_{ij} \pi^{\kappa}_{jn'} \biggl( \frac{1}{E_{ni} E_{nj}} +
   \frac{1}{E_{n'i} E_{n'j}} \biggr) \\ - \frac12 \sum_{i}^{B}
   \sum_{n''}^{A} \biggl( \frac{\pi^{\lambda}_{nn''} \pi^{\mu}_{n''i}
   \pi^{\kappa}_{in'} }{E_{ni} E_{n''i}} + \frac{\pi^{\lambda}_{ni}
   \pi^{\mu}_{in''} \pi^{\kappa}_{n''n'}}{E_{n'i} E_{n''i}} \biggr) ,
   \label{eq:C_total}
\end{multline}
\begin{multline}
   Q^{\lambda\mu\kappa\nu}_{nn'} =
   \tilde{Q}^{\lambda\mu\kappa\nu}_{nn'} + \frac12 \sum_{i}^{B}
   (\tilde{D}^{\lambda\mu}_{ni} \tilde{D}^{\kappa\nu}_{in'} +
   \pi^{\lambda}_{ni} \tilde{C}^{\mu\kappa\nu}_{in'} +
   \tilde{C}^{\lambda\mu\kappa}_{ni} \pi^{\nu}_{in'}) \biggl(
   \frac{1}{E_{ni}} + \frac{1}{E_{n'i}} \biggr) \\ + \frac12
   \sum_{i}^{B} \sum_{j}^{B} ( \tilde{D}^{\lambda\mu}_{ni}
   \pi^{\kappa}_{ij} \pi^{\nu}_{jn'} + \pi^{\lambda}_{ni}
   \tilde{D}^{\mu\kappa}_{ij} \pi^{\nu}_{jn'} + \pi^{\lambda}_{ni}
   \pi^{\mu}_{ij} \tilde{D}^{\kappa\nu}_{jn'} ) \biggl( \frac{1}{E_{ni}
   E_{nj}} + \frac{1}{E_{n'i} E_{n'j}} \biggr) \\ - \frac12
   \sum_{i}^{B} \sum_{n''}^{A} \biggl(
   \frac{\tilde{D}^{\lambda\mu}_{nn''} \pi^{\kappa}_{n''i}
   \pi^{\nu}_{in'} + \pi^{\lambda}_{nn''} \tilde{D}^{\mu\kappa}_{n''i}
   \pi^{\nu}_{in'} + \pi^{\lambda}_{nn''} \pi^{\mu}_{n''i}
   \tilde{D}^{\kappa\nu}_{in'}}{E_{ni} E_{n''i}} +
   \frac{\tilde{D}^{\lambda\mu}_{ni} \pi^{\kappa}_{in''}
   \pi^{\nu}_{n''n'} + \pi^{\lambda}_{ni} \tilde{D}^{\mu\kappa}_{in''}
   \pi^{\nu}_{n''n'} + \pi^{\lambda}_{ni} \pi^{\mu}_{in''}
   \tilde{D}^{\kappa\nu}_{n''n'}}{E_{n'i} E_{n''i}} \biggr) \\ +
   \order (\pi^4) . \label{eq:Xi_total}
\end{multline}
In the last expression, the symbol $\order (\pi^4)$ denotes terms of
the fourth order in the kinetic momentum $\pi$.  These cannot be
obtained from Eq.\ (\ref{eq:bp138}), but they can be derived easily
from Eq.\ (B15e) on p.~205 of Ref.\ \onlinecite{Wink03}.

The term $W^{(2)}_{nn'} (\vect{q})$ is a renormalized second-order
contribution to the band offsets, given by
\begin{equation}
   W^{(2)}_{nn'} (\vect{x}) = \tilde{W}^{(2)}_{nn'} (\vect{x}) +
   \frac12 \sum_{i}^{B} \biggl( \frac{1}{E_{ni}} + \frac{1}{E_{n'i}}
   \biggr) [W^{(1)}_{ni} (\vect{x}) + \Lambda_{ni} \varphi^{(1)}
   (\vect{x})] [W^{(1)}_{in'} (\vect{x}) + \Lambda_{in'} \varphi^{(1)}
   (\vect{x})] , \label{eq:W2_total}
\end{equation}
where $W^{(1)}_{nn'} (\vect{q})$ and $\tilde{W}^{(2)}_{nn'}
(\vect{q})$ were defined in Eqs.\ (\ref{eq:W1kk}) and (\ref{eq:W2b}).
The only contribution from $\varphi^{(1)}$ to be included here is the
linear interface dipole term (\ref{eq:n1_Taylor}) for heterovalent
systems; but note that this contribution is zero in the $GW$
approximation \cite{Hedin65,HedLund69} and in density-functional
theory \cite{HoKo64,KoSh65} (since $\Lambda_{nn'} = \delta_{nn'}$ in
these cases).

The remaining terms in (\ref{eq:em1}) are renormalized versions of the
functions $\tilde{J}^{\lambda}_{nn'} (\vect{q})$,
$\tilde{M}^{\lambda\mu}_{nn'} (\vect{q})$, and
$\tilde{R}^{\lambda\mu}_{nn'} (\vect{q})$ defined in Eqs.\
(\ref{eq:W1kk}) and (\ref{eq:Rq}):
\begin{equation}
   J^{\lambda}_{nn'} (\vect{q}) = \tilde{J}^{\lambda}_{nn'} (\vect{q})
   + \frac12 \sum_{i}^{B} \pi^{\lambda}_{ni} W^{(1)}_{in'} (\vect{q})
   \biggl( \frac{1}{E_{ni}} + \frac{1}{E_{n'i}} \biggr) ,
   \label{eq:J_total}
\end{equation}
\begin{multline}
   M^{\lambda\mu}_{nn'} (\vect{q}) = \tilde{M}^{\lambda\mu}_{nn'}
   (\vect{q}) + \frac12 \sum_{i}^{B} \tilde{D}^{\lambda\mu}_{ni}
   W^{(1)}_{in'} (\vect{q}) \biggl( \frac{1}{E_{ni}} +
   \frac{1}{E_{n'i}} \biggr) + \frac12 \sum_{i}^{B} \pi^{\lambda}_{ni}
   \tilde{J}^{\mu}_{in'} (\vect{q}) \biggl( \frac{1}{E_{ni}} +
   \frac{1}{E_{n'i}} \biggr) \\ + \frac12 \sum_{i}^{B} \sum_{j}^{B}
   \pi^{\lambda}_{ni} \pi^{\mu}_{ij} W^{(1)}_{jn'} (\vect{q}) \biggl(
   \frac{1}{E_{ni} E_{nj}} + \frac{1}{E_{n'i} E_{n'j}} \biggr) \\ -
   \frac12 \sum_{i}^{B} \sum_{n''}^{A} \biggl(
   \frac{\pi^{\lambda}_{nn''} \pi^{\mu}_{n''i} W^{(1)}_{in'}
   (\vect{q})}{E_{ni} E_{n''i}} + \frac{\pi^{\lambda}_{ni}
   \pi^{\mu}_{in''} W^{(1)}_{n''n'} (\vect{q})}{E_{n'i} E_{n''i}}
   \biggr) , \label{eq:M_total}
\end{multline}
\begin{multline}
   R^{\lambda\mu}_{nn'} (\vect{q}) = \tilde{R}^{\lambda\mu}_{nn'}
   (\vect{q}) + \frac12 \sum_{i}^{B} \biggl( \pi^{\lambda}_{ni}
   [\tilde{J}^{\mu}_{n'i} (-\vect{q})]^* + \tilde{J}^{\lambda}_{ni}
   (\vect{q}) \pi^{\mu}_{in'} \biggr) \biggl( \frac{1}{E_{ni}} +
   \frac{1}{E_{n'i}} \biggr) \\ + \frac12 \sum_{i}^{B} \sum_{j}^{B}
   \pi^{\lambda}_{ni} W^{(1)}_{ij} (\vect{q}) \pi^{\mu}_{jn'}
   \biggl( \frac{1}{E_{ni} E_{nj}} + \frac{1}{E_{n'i} E_{n'j}} \biggr)
   \\ - \frac12 \sum_{i}^{B} \sum_{n''}^{A} \biggl(
   \frac{\pi^{\lambda}_{nn''} W^{(1)}_{n''i} (\vect{q})
   \pi^{\mu}_{in'} }{E_{ni} E_{n''i}} + \frac{\pi^{\lambda}_{ni}
   W^{(1)}_{in''} (\vect{q}) \pi^{\mu}_{n''n'} }{E_{n'i}
   E_{n''i}} \biggr) . \label{eq:R_total}
\end{multline}
Note that $M^{\lambda\mu}_{nn'} (\vect{q}) \neq M^{\mu\lambda}_{nn'}
(\vect{q})$, in contrast to Eq.\ (\ref{eq:Psym}).

\end{widetext}


\begin{thebibliography}{175}
\bibitem[{not({\natexlab{a}})}]{note:heterovalent}
\bibinfo{note}{There are three possible ways of dealing with heterovalent
  systems: (i)~Treating the monopole terms directly, at the level of the
  approximation scheme defined in Ref.\ \onlinecite{Fore05a}. This is the same
  level of accuracy as ordinary effective-mass theory, \cite{LuKo55} which
  means that the position dependence of the effective mass
  (\ref{eq:pos_dep_mass}) and the interface terms (\ref{eq:d0}) and
  (\ref{eq:d1}) are neglected. (ii)~Treating the monopole terms directly, but
  extending the analysis in Ref.\ \onlinecite{Fore05a} to higher order,
  including nonanalytic terms in $W^{(2)}$ and $n^{(2)}$. (iii)~Grouping the
  monopole terms into neutral clusters of fractional atoms, \cite{Fore05a}
  which are then treated as the fundamental unit of perturbation. This is the
  approach adopted here, because it allows heterovalent and isovalent systems
  to be treated as formally the same. However, since the cluster size is
  several nearest-neighbor spacings, this approach places more stringent limits
  on what constitutes a ``slowly varying'' envelope.}
\bibitem[{not({\natexlab{b}})}]{note:any}
\bibinfo{note}{This statement refers, of course, only to the class of
  heterostructures for which the slowly varying envelope approximation and the
  error analysis in Sec.\ \ref{sec:error} remain valid.}
\bibitem[{not({\natexlab{c}})}]{note:atom}
\bibinfo{note}{Although the language used here (and throughout this paper)
  refers to atoms, in heterovalent systems it is convenient
  \cite{note:heterovalent} to use these symbols as labels for the neutral
  clusters of (fractional) atoms defined in Ref.\ \onlinecite{Fore05a}.}
\bibitem[{not({\natexlab{d}})}]{note:new_phi}
\bibinfo{note}{In Eq.\ (\ref{eq:phi1}), an analytic contribution to
  $\varphi^{\alpha \bm{\tau}}_{\vect{R}} (\vect{k} - \vect{k}')$ from Ref.\
  \onlinecite{Fore05a} has been absorbed into the definition of $w^{\alpha
  \bm{\tau}}_{\vect{R}} (\vect{k}, \vect{k}'; \vect{G}, \vect{G}')$.}
\bibitem[{not({\natexlab{e}})}]{note:nonorthogonal}
\bibinfo{note}{This choice of basis was criticized in Ref.\
  \onlinecite{TakVol00a} as being nonorthogonal and overcomplete. However,
  since (for example) the $X_z$ states change sign under any fcc lattice
  translation whose $z$ component is $\pm a / 2$, the basis functions are
  orthogonal with respect to the simple cubic unit cell. Also, although there
  are four times as many zone-center basis functions in the supercell, a
  primitive unit cell in the reciprocal lattice has a volume $\Omega_0^* =
  (2\pi)^3 / \Omega_0$ only one-fourth the original volume, so the basis is
  complete but not overcomplete.}
\bibitem[{not({\natexlab{f}})}]{note:umklapp}
\bibinfo{note}{It was shown in Refs.\ \onlinecite{CuVH91} and
  \onlinecite{TakVol99b} that when $\Omega_0^*$ is chosen to be the
  conventional 3D Brillouin zone of the reference crystal, difficulties
  involving ``umklapp''-like processes generated by $V_{nn'}(\vect{k},
  \vect{k}')$ may arise in certain applications for heterostructures with 2D
  translation symmetry. However, as noted by Burt, \cite{Burt92} these
  difficulties are easily avoided by choosing a ``slab-adapted'' unit cell.
  \cite{Heine63,Klein81} It is assumed here that this choice has been made.}
\bibitem[{not({\natexlab{g}})}]{note:symmetric}
\bibinfo{note}{The expression given by Leibler in Eq.\ (30) of Ref.\
  \onlinecite{Leib77} is antisymmetric in the Cartesian indices, but this is
  obviously incorrect, as the net contribution from this term would then be
  identically zero.}
\bibitem[{not({\natexlab{h}})}]{note:Vmodel}
\bibinfo{note}{This potential was of the form (\ref{eq:Vmodel}), but with
  $\theta^{l} (\vect{q}) = 0$ outside the first Brillouin zone.}
\end{thebibliography}

\end{document}